\documentclass[aps,twocolumn,superscriptaddress,floatfix]{revtex4-2}

\usepackage{amsmath,amssymb}
\usepackage{graphicx}
\usepackage{hyperref}
\usepackage{xcolor}
\usepackage{bm}
\usepackage{booktabs}

\begin{document}

\title{Microscopic dynamics of consensus formation in multi-agent LLM Naming Games}

\author{Cristiano De Nobili}
\altaffiliation[email: ]{cristiano@critiqality.ai}
\affiliation{Critiqality, Via Pinturicchio 21, 20133, Milan, Italy}
\author{Vijayasri Iyer}
\affiliation{Independent Researcher}
\author{Alessandro Codello}
\affiliation{DSMN, Ca'\ Foscari University of Venice, Via Torino 155, 30172 Venice, Italy}
\affiliation{IFFI, Universidad de la Rep\'ublica, J.H.y Reissig 565, 11300 Montevideo, Uruguay}
\author{Raffaella Burioni}
\affiliation{Dipartimento di Scienze Matematiche, Fisiche e Informatiche,
Universit\`a degli Studi di Parma, Parco Area delle Scienze, 7/A 43124 Parma, Italy}
\affiliation{INFN, Gruppo Collegato di Parma, Parco Area delle Scienze 7/A, 43124 Parma, Italy}


\begin{abstract}
Decentralized populations of Large Language Model (LLM) agents can spontaneously reach consensus on shared conventions, yet the microscopic mechanisms by which their internal stochasticity shapes macroscopic ordering remain unexplored. We study a minimal LLM Naming Game in which the listener's decision is a single-token LLM call at decoding temperature $T$, replacing the inventory check of the deterministic Naming Game. Each interaction decomposes into an in-inventory and an out-inventory channel with conditional rates $\pi(T)\!\equiv\!P(\text{YES}\mid w\in P_j)$ and $\phi(T)\!\equiv\!P(\text{YES}\mid w\notin P_j)$, whose balance controls an ordering-disordering drift. A mean-field theory of the two-rate dynamics yields an analytical ordering condition that generalizes the consensus threshold of the stochastic Naming Game to a critical line in the $(\pi,\phi)$ plane. Across three open-weight architectures, consensus is always reached, but through three distinct listener regimes: permissive (repaint-noise dominated), near-deterministic, and conservative (missed-collapse dominated). The effective finite-size exponent $\beta(T)$ in $t_{\rm conv}\!\sim\!N^{\beta}$ shifts with temperature, and the temperature-sensitivity $\alpha$ in $t_c\!\sim\!e^{\alpha T}$ ranges from ${\approx}\,0.67$ to ${\approx}\,0$ across architectures. Decoding temperature thus emerges as an architecture-dependent control parameter for decentralized LLM populations, quantitatively characterized by the statistical-physics toolkit.
\end{abstract}

\maketitle

\section{Introduction}
\label{sec:intro}

Consensus formation, that is the spontaneous emergence of a shared convention
from purely local interactions, is a central problem at the intersection of
statistical physics, complex systems, and multi-agent
intelligence~\cite{castellano2009statistical}. Classical agent-based models
such as the Voter model, Axelrod dynamics, and bounded-confidence frameworks
have shown that macroscopic order can arise from simple microscopic
stochastic rules, with the nature of the resulting phase transitions,
coarsening dynamics, and scaling laws depending sensitively on the
interaction topology and on the agents' decision
rule~\cite{castellano2009statistical}. Among these, the Naming Game
(NG)~\cite{steels1995,baronchelli2006sharp,loreto2011statistical} occupies a
distinguished position: it provides a minimal, exactly solvable model of how
a population bootstraps a shared vocabulary through pairwise negotiation,
exhibiting symmetry breaking and a convergence time
$t_{\rm conv}\sim N^{3/2}$ on fully connected
graphs~\cite{baronchelli2006sharp}. The \emph{deterministic} NG has been studied on
regular lattices~\cite{baronchelli2006topology}, complex
networks~\cite{dall2006nonequilibrium}, and generalized to include
irresolute agents with a tunable commitment parameter, revealing a
non-equilibrium phase transition from consensus to
fragmentation~\cite{baronchelli2007nonequilibrium}.

The rapid deployment of Large Language Models (LLMs) as autonomous,
interacting agents has opened a new arena for these questions. LLM-based
multi-agent systems are now used in negotiation, planning, and
tool-use~\cite{guo2024largelanguage,hong2024metagpt}, and populations of
such agents interact with one another and with humans at increasing
scale~\cite{johnson2026increasing}. Whether and how decentralized
populations of LLMs can self-organize on shared conventions is therefore
not only a foundational scientific question but also a prerequisite for
the safe and reliable deployment of agentic AI. Recent empirical studies
have begun to answer it: Ashery, Aiello, and
Baronchelli~\cite{ashery2025emergent} demonstrated that populations of LLM
agents playing a minimal coordination game, of the type used in human
experiments on convention formation~\cite{centola2015spontaneous}, spontaneously reach
group-wide linguistic conventions and exhibit emergent collective biases
not present at the single-agent level. De Marzo, Castellano, and
Garcia~\cite{demarzo2025aiagentscoordinatehuman} showed that LLM agents
can self-organize on arbitrary binary choices but only below a
model-dependent critical group size, and Flint
\emph{et al.}~\cite{flint2025groupsizeeffectscollective} developed a
mean-field analytical framework relating group size to the structure of
basins of attraction. Related work has investigated conformity and social influence in LLM
agents~\cite{demarzo2026conformitygeneratescollectivemisalignment,bellina2026conformity},
detailed balance in LLM-driven
dynamics~\cite{song2025detailedbalance}, cultural attractors in
transmission chains~\cite{perez2025telephone}, opinion dynamics in
networks of LLMs~\cite{chuang2024wisdom,piatti2024cooperate}, agreement
protocols among reasoning agents~\cite{ruan2025reachingagreementreasoningllm},
and Ising-like collective alignment on
lattices~\cite{denobili2026collectivealignmentllmmultiagent}. Mean-field
methods from statistical mechanics have also proven effective in
large-population multi-agent reinforcement
learning~\cite{yang2018mean}, and differentiable surrogates of
agent-based models are being developed to bridge microscopic rules and
emergent collective phenomena~\cite{cozzi2025learning}.

A consistent picture is emerging: LLM populations \emph{do} reach
consensus on shared conventions, the resulting macroscopic dynamics can
be modelled with statistical-physics tools, and the outcome depends in
non-trivial ways on the underlying architecture. What is still missing,
however, is a microscopic theory linking the stochasticity of LLM
inference to the macroscopic dynamics of consensus. In particular, the
decoding temperature~$T$, the primary hyperparameter governing the
randomness of each agent's output~\cite{holtzman2020curious}, has not been studied as an effective
control parameter in the statistical-physics sense; existing LLM-NG
studies fix~$T$ to a single
value~\cite{ashery2025emergent,flint2025groupsizeeffectscollective}.
This gap is especially relevant as LLM agents move toward decentralized
deployment, where no central coordinator exists and collective outcomes
emerge bottom-up from local
interactions~\cite{johnson2026increasing,cozzi2025learning}.

In this work we fill this gap. We implement the minimal NG with LLM
agents, retaining the original topology and update rule of the \emph{deterministic}
NG but replacing the listener's deterministic inventory check with a
single-token LLM call at temperature~$T$. We then introduce a
\emph{microscopic decomposition} of each interaction into four channels,
parametrized by two conditional rates, $\pi(T)$ and $\phi(T)$, and show
that this pair suffices to organize the model-dependent phenomenology
into three qualitatively distinct regimes. The
setup is deliberately minimal: it isolates the effect of LLM-generated
stochasticity on a well-understood ordering dynamics, free from
prompt engineering, multi-turn memory, or strategic play. Our main
contributions are:
\textit{(i)}~the $(\pi,\phi)$ decomposition itself, which compresses
the stochasticity of LLM inference into two measurable conditional
rates and thereby exposes the microscopic origin of collective order;
\textit{(ii)}~a drift analysis that links these microscopic rates to the net ordering tendency;
\textit{(iii)}~the identification of three qualitatively distinct
listener regimes (permissive, near-deterministic, and conservative) that
emerge from the interplay of architecture and temperature, with
qualitatively different macroscopic signatures including an
\emph{inverted temperature ordering} for the conservative regime;
\textit{(iv)}~finite-size scaling and temperature-response measurements
over the accessible range $N\!\in\![50,150]$ indicating that the effective
exponent $\beta(T)$ in $t_{\rm conv}\!\sim\!N^{\beta}$ varies with decoding
temperature and, for the most noise-dominated architecture, reaches values
above the canonical mean-field $3/2$, and that the exponential rate
$\alpha$ governing $t_c(T)$ is itself architecture-dependent, ranging from
essentially zero to ${\approx}\,0.7$ across the three models tested.
\textit{(v)}~a complete-graph mean-field theory of the two-rate
dynamics whose two-word sector yields the analytical ordering condition
$3\pi-2\phi-1>0$, generalizing the known threshold of the stochastic NG
and rationalizing the three regimes.
Together these results promote decoding temperature to a bona-fide
effective control parameter for decentralized LLM populations, and show
that the statistical-physics toolkit of coarsening, drift analysis, and
scaling provides a natural language for characterizing and ultimately
designing the emergent behaviour of multi-agent AI systems.

\section{Model}
\label{sec:model}
\begin{figure}[t]
  \centering
  \includegraphics[width=0.45\textwidth]{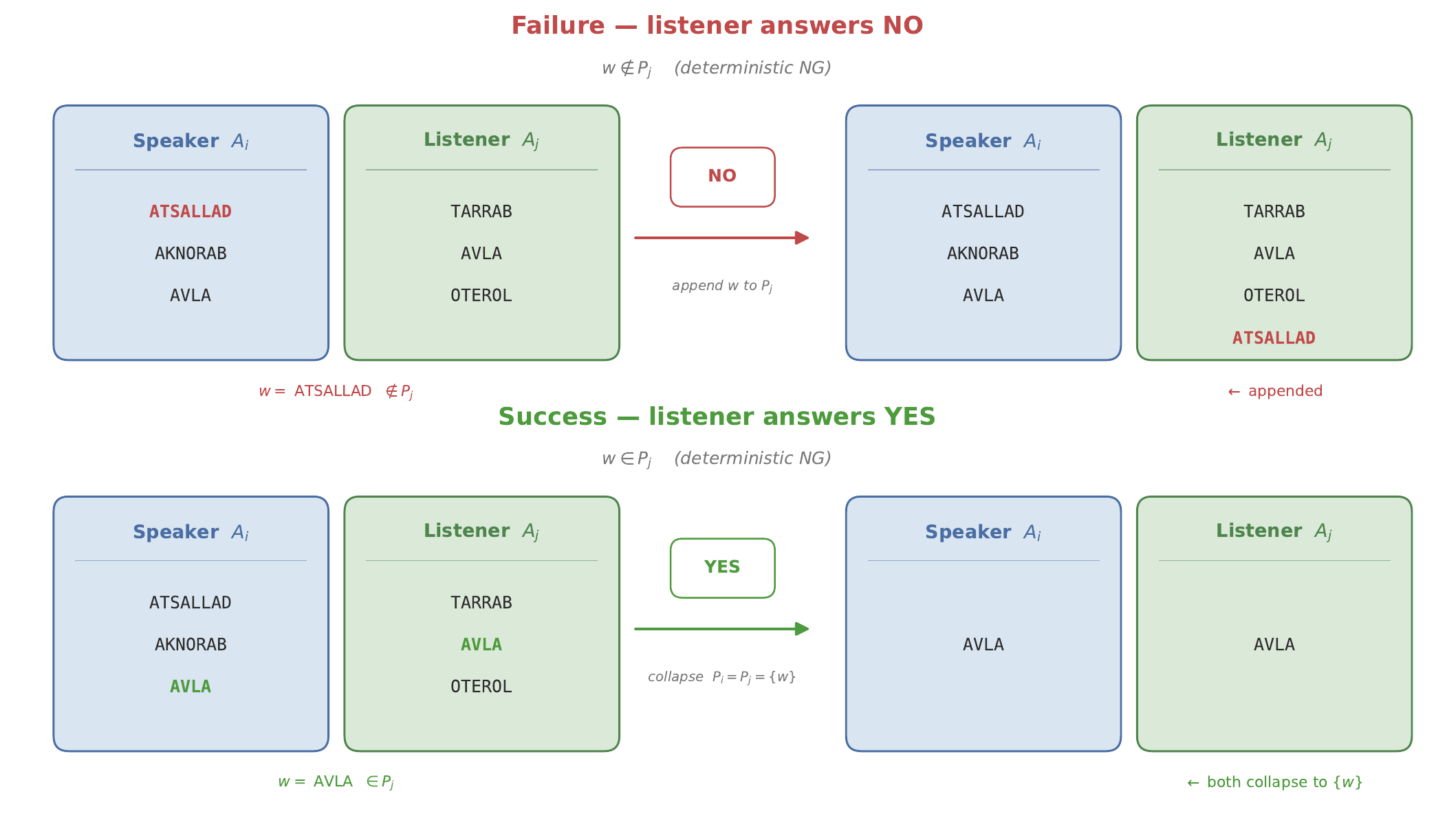}
  \caption{\label{fig:ngrules}%
  Schematic of the Naming Game interaction. In the \emph{deterministic} NG, the decision step is an inventory check ($w \in P_j$); in the LLM-NG it is replaced by the listener's LLM call at temperature~$T$.}
\end{figure}

In the \emph{deterministic NG}, $N$ agents are placed on a fully connected graph. Each agent~$i$ carries an inventory $P_i(t)\subseteq\mathcal{V}$, where $\mathcal{V}$ is a pool of words. All inventories start empty. At each step~$t$, an ordered pair $(i,j)$ is drawn uniformly (speaker~$i$, listener~$j$). If an agent has an empty inventory it invents by sampling from $\mathcal{V}$. The speaker picks $w$ uniformly from~$P_i$ and transmits it. The interaction succeeds if $w\in P_j$: both agents collapse to $\{w\}$; otherwise $j$ appends~$w$ (Fig.~\ref{fig:ngrules}). On the fully connected graph, the \emph{deterministic} NG exhibits a characteristic three-stage dynamics: an initial phase in which agents
accumulate distinct words in their inventories, a coarsening stage in
which pairwise agreements build up correlations between inventories,
and a final rapid collapse to a single conventional
name~\cite{baronchelli2006sharp,loreto2011statistical}. The
resulting consensus time scales as $t_c\sim N^{3/2}$, a benchmark
against which we compare the LLM-driven dynamics throughout this
work.

\begin{figure}[t]
  \centering
  \includegraphics[width=0.45\textwidth]{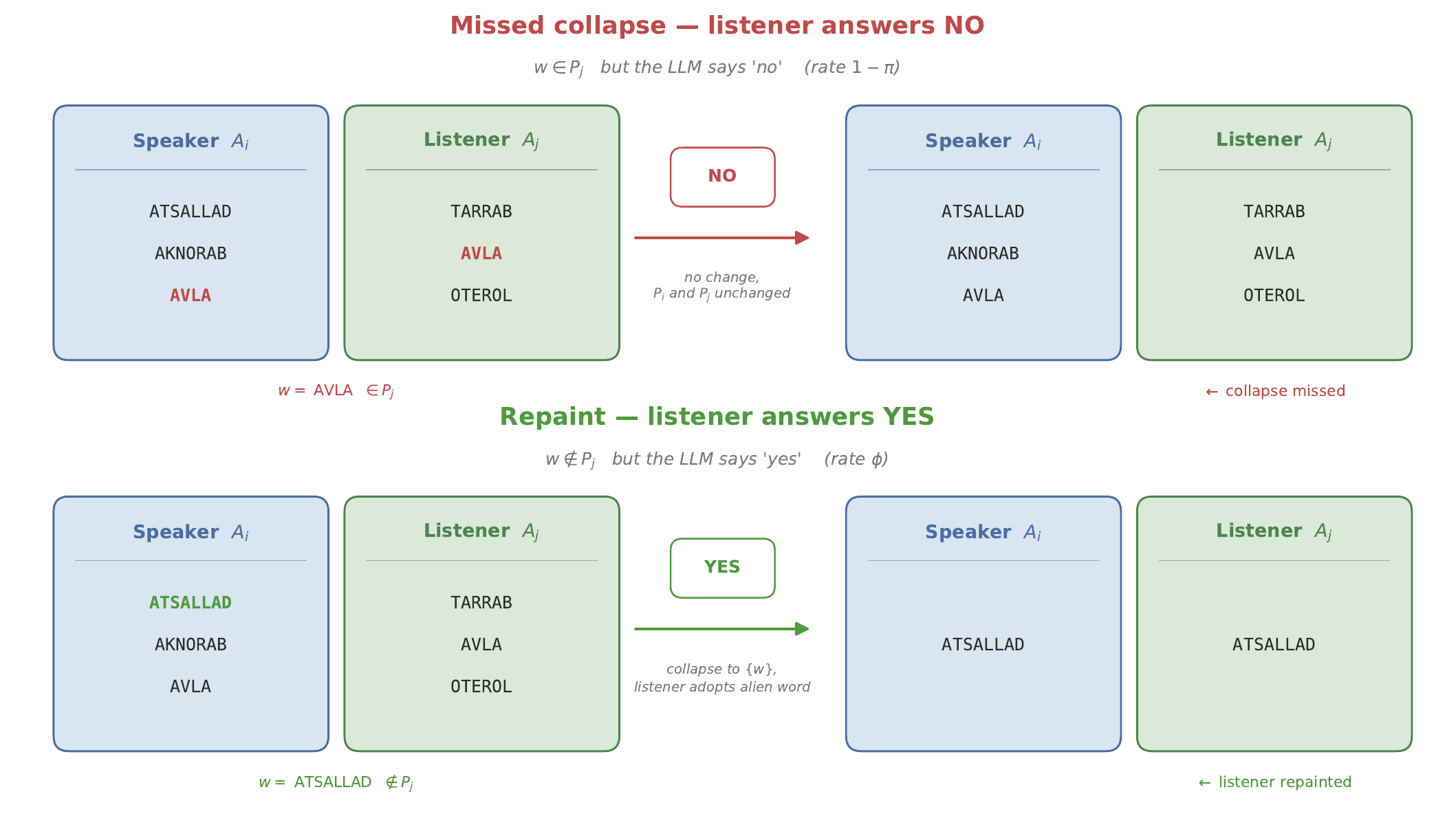}
  \caption{\label{fig:llmngrules}%
  The two new interaction channels of the LLM-NG that do not exist in
  the deterministic NG (Fig.~\ref{fig:ngrules}). Top: a \emph{missed
  collapse} occurs when $w\!\in\!P_j$ but the LLM answers NO, at rate
  $1{-}\pi$; the inventories are left unchanged. Bottom: a
  \emph{repaint} occurs when $w\!\notin\!P_j$ but the LLM answers YES,
  at rate $\phi$; the listener discards its inventory and adopts the
  alien word~$w$. Together with Fig.~\ref{fig:ngrules}, these panels
  enumerate the four channels (TP, FN, FP, TN) into which the LLM-NG
  interaction decomposes (Sec.~\ref{sec:micro}).}
\end{figure}

In the \emph{LLM-NG}, the listener receives a prompt containing its inventory and the proposed word, and returns a single YES/NO token at temperature~$T$. If YES, both collapse to $\{w\}$; if NO, $j$ appends~$w$. The prompt used throughout our simulations is:
\texttt{System:} ``\textit{You are an agent with your own language and vocabulary. You can and must reply with yes or no.}''
\texttt{User:} ``\textit{Your words are: $P_j$. Do we add $w$ to the
list?}''. The only control parameter varied is~$T$. In this work, we chose $|\mathcal{V}|\!=\!10^4$ English words. This number is much larger than any explored population size, so inventory collisions during invention are negligible and $\mathcal{V}$ is effectively unlimited, as in the \emph{deterministic} NG~\cite{baronchelli2006sharp}. We track: the number of distinct words $N_d(t)=|\bigcup_i P_i(t)|$ and the success rate $S(t)$. Consensus is reached at $t_c$ defined by $N_d(t_c)=1$. 

A comment on the prompt. Any wording choice can induce a listener bias beyond the architecture- and temperature-dependent one under study. Alternative phrasings such as ``Is the proposed word already in your list? Answer only YES or NO.'' are equally legitimate and could shift the numerical values of $(\pi,\phi)$. A parallel statistical-physics study of LLM populations on 2D lattices~\cite{denobili2026collectivealignmentllmmultiagent} observes that prompt rewordings can move the effective microscopic parameters quantitatively while leaving the overall collective behaviour and its physical description qualitatively similar. Given the compute budget of the present study we did not perform a systematic prompt-variation scan and defer this robustness check to future work.

\section{Microscopic decomposition}
\label{sec:micro}
At each interaction, the proposed word $w$ is either already present in the listener's inventory, or absent from it. We refer to these as the in-inventory and out-inventory channels, respectively. The LLM listener can answer YES or NO in either case, producing four microscopic outcomes: 
true positive (TP: $w\in P_j$, YES),
false negative (FN: $w\in P_j$, NO),
false positive (FP: $w\notin P_j$, YES),
and true negative (TN: $w\notin P_j$, NO).
We define two conditional acceptance rates (Fig.~\ref{fig:llmngrules}):
\begin{align}
  \pi(T) &\equiv P(\text{YES}\mid w\in P_j) = \frac{\text{TP}}{\text{TP}+\text{FN}}\,,
  \label{eq:pi}\\
  \phi(T) &\equiv P(\text{YES}\mid w\notin P_j) = \frac{\text{FP}}{\text{FP}+\text{TN}}\,.
  \label{eq:phi}
\end{align}
The \emph{deterministic} NG corresponds to $\pi\!=\!1$, $\phi\!=\!0$: $\pi(T)$
is the rate at which a collapse is correctly triggered, $\phi(T)$ the
rate at which one is triggered erroneously. To weight these rates by how
often each situation occurs, we introduce the \emph{in-inventory fraction}
\begin{equation}
  m(t) \;\equiv\; P\bigl(w\!\in\!P_j(t)\bigr)\,,
  \label{eq:m}
\end{equation}
that is, the probability that at time~$t$ the transmitted word~$w$ already
belongs to the listener's inventory, averaged over uniformly random
speaker--listener pairs $(i,j)$ and over words $w$ drawn uniformly from
$P_i(t)$. Empirically, $m(t)$ is estimated from the running fraction of in-inventory interactions,
\begin{equation}
  \hat{m}(t) \;=\; \frac{\mathrm{TP}(t)+\mathrm{FN}(t)}{\mathrm{TP}(t)+\mathrm{FN}(t)+\mathrm{FP}(t)+\mathrm{TN}(t)}\,,
  \label{eq:mhat}
\end{equation}
evaluated in a small window around~$t$. By construction $m(t)$ starts near
zero (empty or disparate inventories) and approaches unity at consensus.
The two rates then control competing tendencies: an in-inventory--YES event
(rate $m\,\pi$) is a legitimate collapse, the ordering channel; an
out-inventory--YES event (rate $(1{-}m)\,\phi$) is a \emph{repaint}, in which
the listener discards its inventory and adopts an alien word, the
disordering channel. Two clarifications on this labelling are in order. First,
``disordering'' is an ensemble-averaged statement, not a property of every
event: a single repaint does commit both interacting agents to the
singleton $\{w\}$, but because $w$ is by construction \emph{not} the word
the listener held, repeated repaints on average redirect the population
toward random attractors rather than toward the name emerging as the
global consensus. Second, repaints are not always harmful: near
consensus, when only two or three names coexist in a metastable
configuration, a rare repaint can break the deadlock by knocking an agent
out of one attractor so that the majority absorbs it. This is how a small
residual~$\phi$ can accelerate late-time convergence, and why the drift
proxy introduced below predicts the sign of ordering only on average. In
what follows, ``ordering'' and ``disordering'' should be read as ensemble
labels valid at first order in a mean-field description.

We define a drift proxy
\begin{equation}
  \Delta(t) = m(t)\,\pi(t) - \lambda\,(1-m(t))\,\phi(t)\,,
  \label{eq:drift}
\end{equation}
with $\lambda=1$. Positive drift implies net ordering; negative drift signals
that repaint noise overwhelms consolidation. The choice $\lambda\!=\!1$ deserves a brief comment. A
true-positive collapse consolidates two agents around a word they already
share, leaving the number of $w$-holders unchanged; a false-positive
collapse instead recruits the listener into the population of $w$-holders from scratch, forcing it to abandon its entire prior inventory. The consequences for
$N_d$, for individual inventory sizes, and for the overlap distribution
therefore differ in magnitude between the two events, and a rigorous
mean-field treatment would yield $\lambda\neq 1$, possibly state-dependent.
We adopt $\lambda\!=\!1$ as the simplest phenomenological choice: since it
is the \emph{sign} of $\Delta$, and not its magnitude, that determines the
ordering versus disordering balance and hence the three regimes of
Sec.~\ref{sec:rates}, this suffices for the qualitative claims made here. A
mean-field derivation of $\lambda$ from the microscopic rates, along the
lines of Ref.~\cite{baronchelli2007nonequilibrium}, is a natural
extension.

The $(\pi,\phi)$ plane
organizes several known limits: the \emph{deterministic} NG
($\pi\!=\!1,\phi\!=\!0$)~\cite{baronchelli2006sharp}, lazy consolidation
($\pi\!<\!1,\phi\!=\!0$), voter-like ($\pi\!=\!1,\phi\!=\!1$), and
maximal chaos ($\pi\!=\!0,\phi\!=\!1$). The lazy-consolidation edge is
the stochastic negotiation model of Baronchelli, Dall'Asta, Barrat, and
Loreto~\cite{baronchelli2007nonequilibrium}, whose hand-tuned
commitment probability~$\beta$ plays the role of $\pi$, with $\phi$
identically zero by construction because out-inventory interactions cannot
trigger a collapse in their update rule. The decisive difference, developed
in Sec.~\ref{sec:discussion}, is that their $\beta$ is imposed externally on
the update rule, whereas our $\pi(T)$ and $\phi(T)$ are emergent outputs of
the LLM listener, measured a posteriori from actual multi-agent
simulations.

\section{Results}
\label{sec:results}

We simulate the LLM-NG for three open-weight models served locally via
Ollama~\cite{ollama}: \texttt{llama3.1:8b} (Meta),
\texttt{mistral:7b} (Mistral~AI), and \texttt{phi3:14b} (Microsoft).
Unless stated otherwise, $N=150$ agents interact for up to $10^5$ steps
($1.75\times 10^5$ for \texttt{phi3:14b}), with
$T\in\{0.05,0.2,0.4,0.6,0.8,1.0,1.2,1.4,1.6,1.8,2.0\}$.
Each configuration is averaged over $10$ to $15$ seeds.
Throughout, consensus is defined by strict $1$-consensus
($N_d(t_c)\!=\!1$); central lines in all consensus-time plots show the
\emph{median} across seeds, and shaded bands span the interquartile range
(p25, p75). This robust statistic was preferred over the mean and standard
deviation because of the heavy-tailed seed distributions characteristic of
the conservative listener regime.

\subsection{Microscopic rates}
\label{sec:rates}

\begin{figure*}[t]
  \centering
  \includegraphics[width=0.325\textwidth]{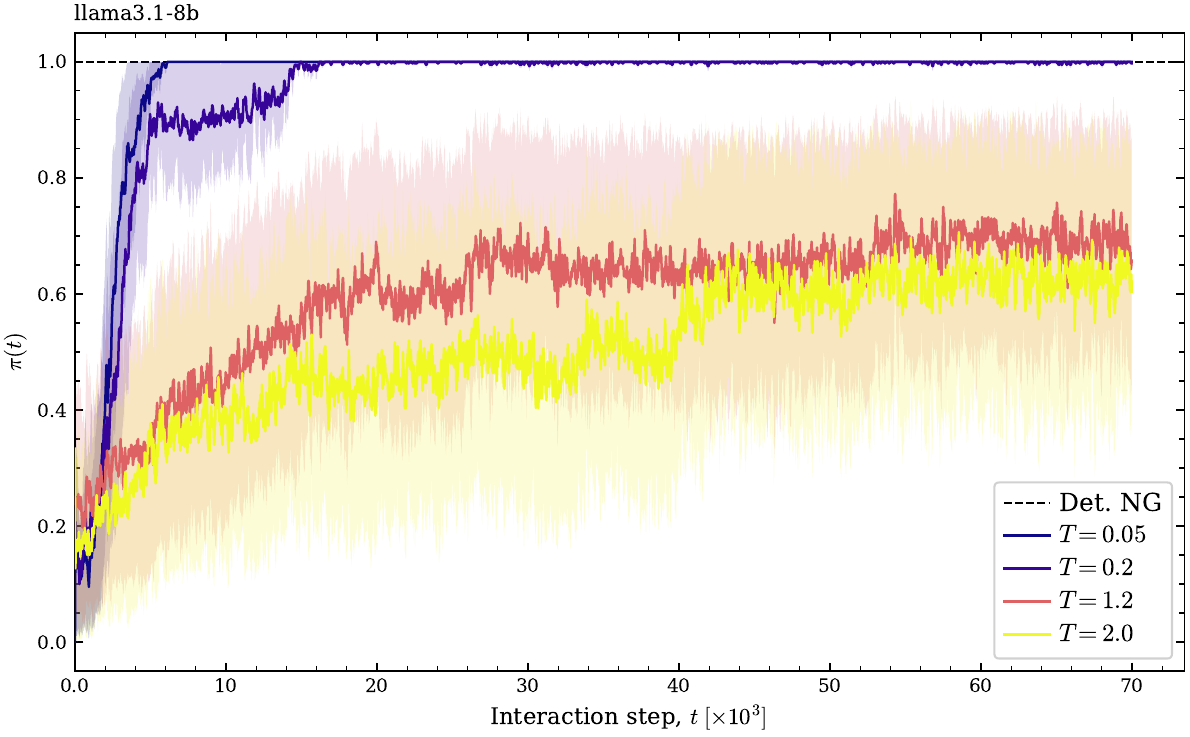}%
  \hfill
  \includegraphics[width=0.325\textwidth]{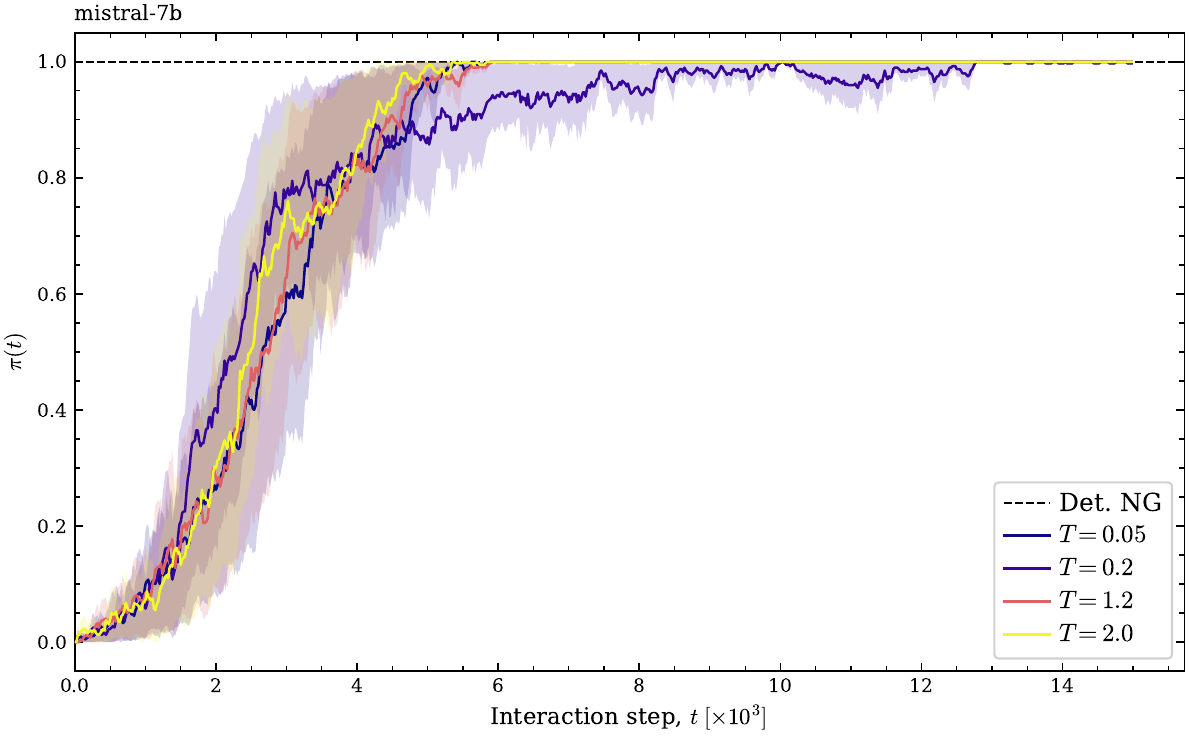}%
  \hfill
  \includegraphics[width=0.325\textwidth]{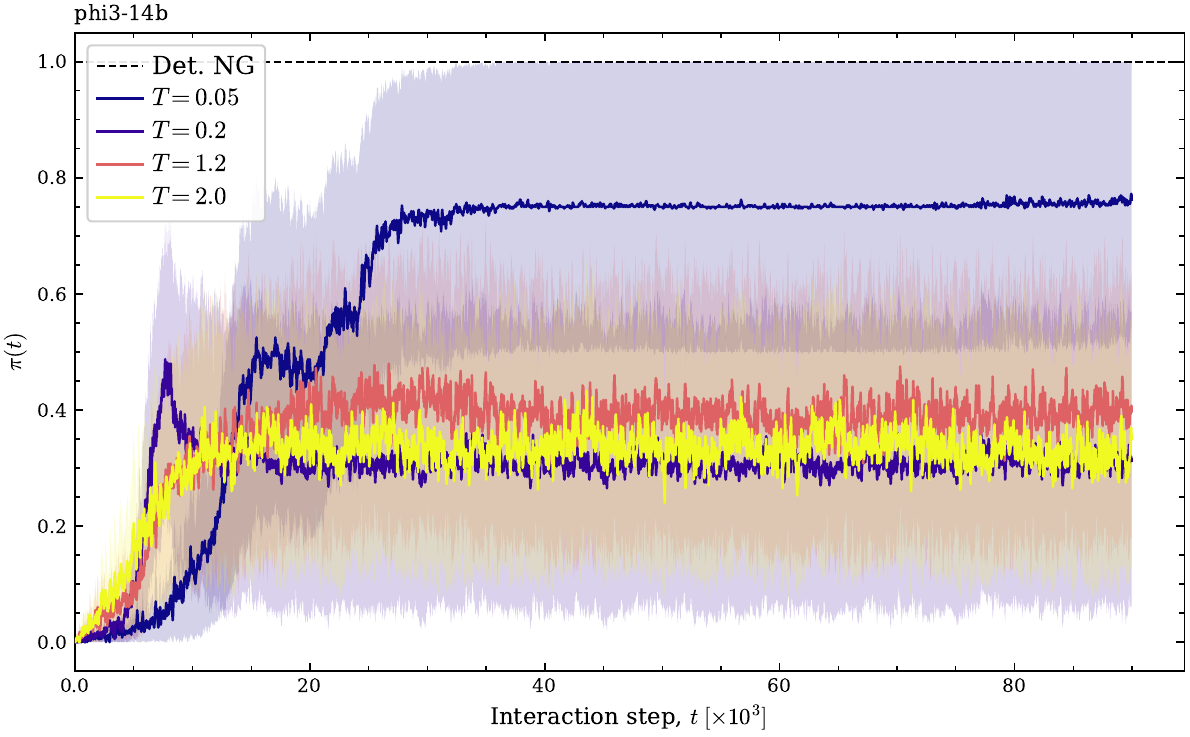}\\[4pt]
  \includegraphics[width=0.325\textwidth]{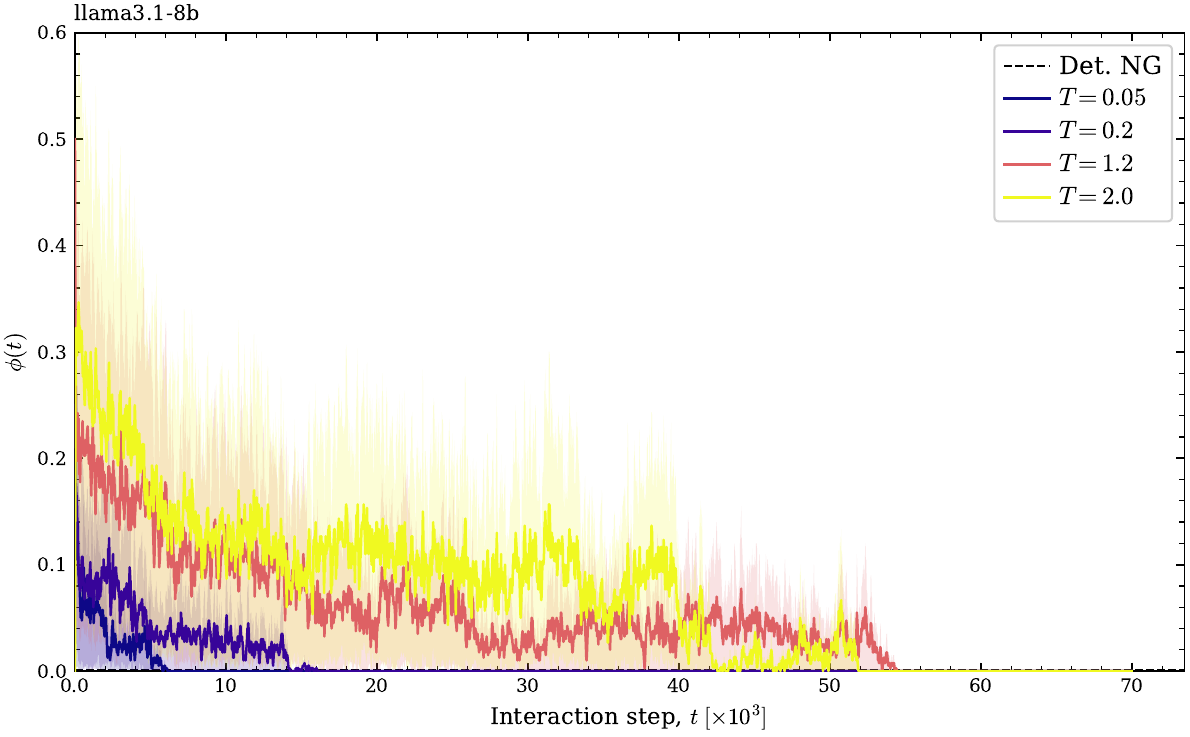}%
  \hfill
  \includegraphics[width=0.325\textwidth]{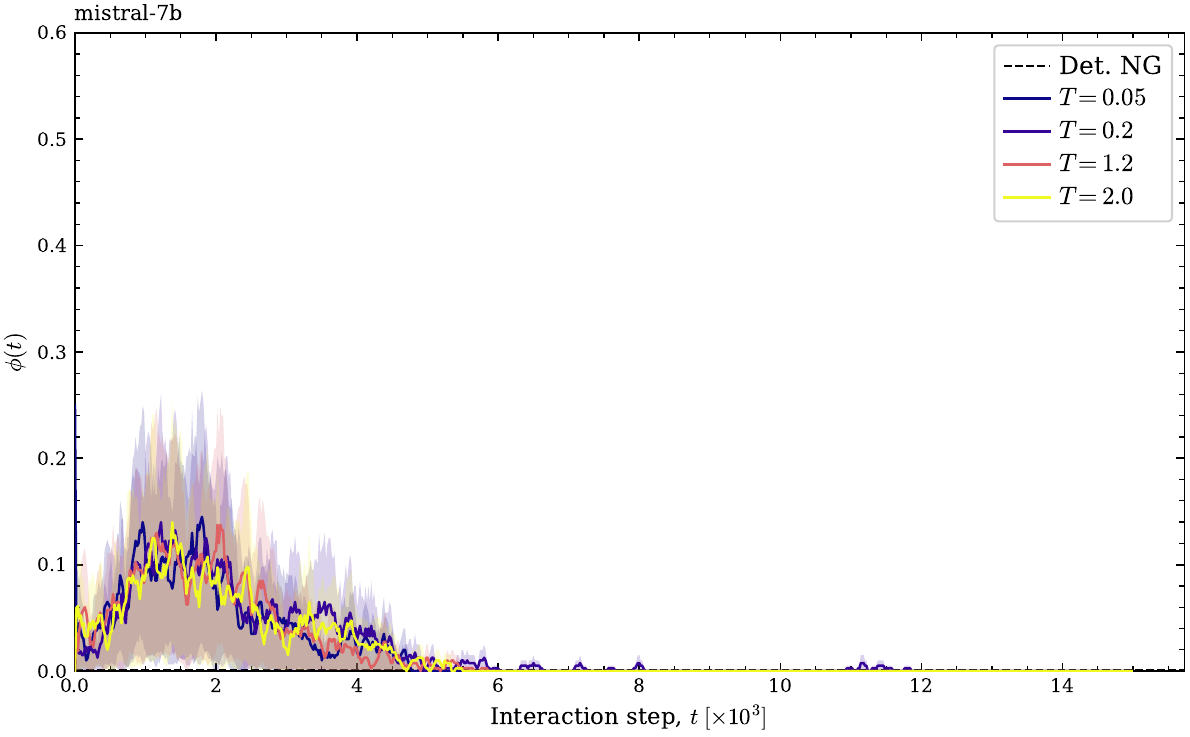}%
  \hfill
  \includegraphics[width=0.325\textwidth]{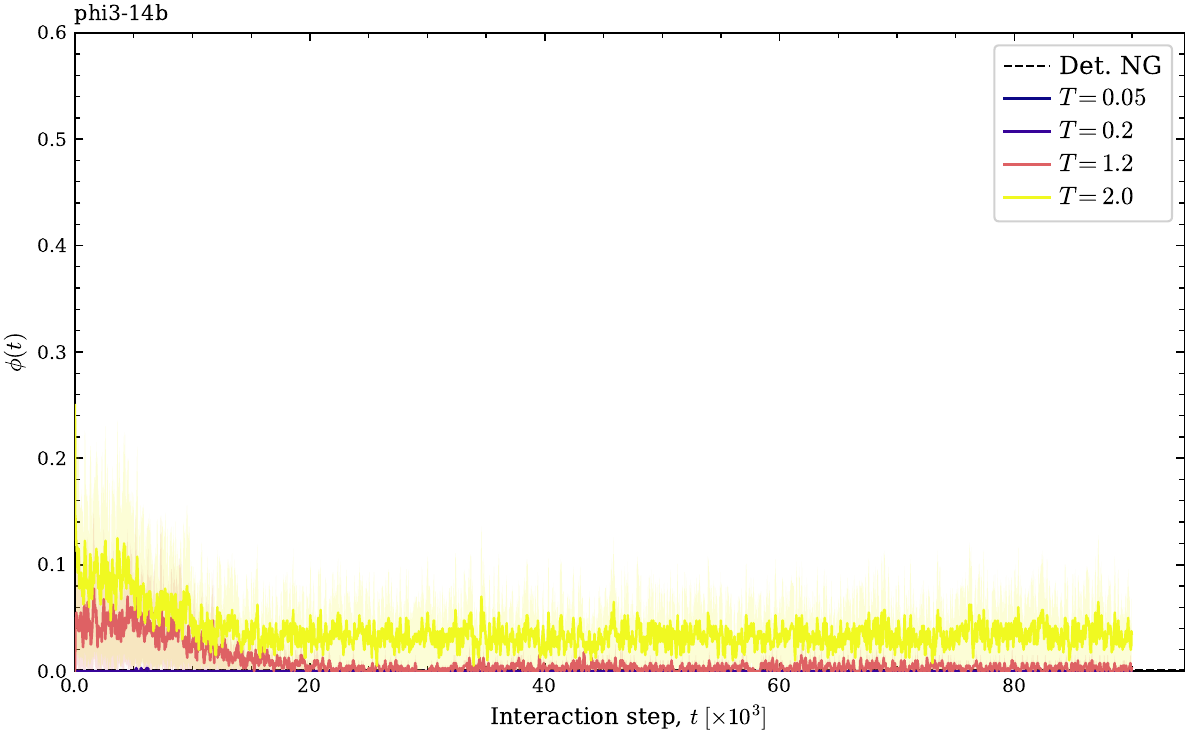}
  \caption{\label{fig:piandphi}%
  Microscopic conditional rates for $N\!=\!150$ agents at selected
  temperatures. \textbf{Top row:} consolidation rate
  $\pi(t)\!=\!P(\text{YES}\mid w\in P_j)$ for
  (a)~\texttt{llama3.1:8b}, (b)~\texttt{mistral:7b}, (c)~\texttt{phi3:14b}.
  Dashed line: \emph{deterministic} NG ($\pi\!=\!1$). \textbf{Bottom row:} repaint
  rate $\phi(t)\!=\!P(\text{YES}\mid w\notin P_j)$ for
  (d)~\texttt{llama3.1:8b}, (e)~\texttt{mistral:7b}, (f)~\texttt{phi3:14b}.
  Dashed line: \emph{deterministic} NG ($\phi\!=\!0$). Shaded bands indicate
  seed-to-seed variance. Note the different $y$-axis scales in the bottom
  row, reflecting the order-of-magnitude difference in repaint noise across
  models.}
\end{figure*}

Figure~\ref{fig:piandphi} summarizes the microscopic behaviour of the three
models through the conditional rates $\pi(t)$ and $\phi(t)$.

\texttt{llama3.1:8b} \emph{(permissive listener).} The consolidation rate
$\pi(t)$ settles at a temperature-dependent plateau: ${\approx}\,1.0$ at
$T\!=\!0.05$, decreasing to ${\approx}\,0.55$ at $T\!=\!2.0$
(Fig.~\ref{fig:piandphi}a). The repaint rate $\phi(t)$ starts high and
decays as the system orders (Fig.~\ref{fig:piandphi}d): at $T\!=\!2.0$,
$\phi$ reaches ${\approx}\,0.50$ at early times, remaining elevated for
tens of thousands of steps. Both $\pi$ and $\phi$ show clear, monotonic
temperature ordering. This places llama in a \emph{repaint-noise dominated}
regime where $\phi$ is the main lever through which temperature controls
the dynamics.

\texttt{mistral:7b} \emph{(near-deterministic listener).} $\pi(t)$
saturates at ${\approx}\,1.0$ for \emph{all} tested temperatures
(Fig.~\ref{fig:piandphi}b), and $\phi(t)$ shows only a small early-time
hump (${\lesssim}\,0.15$) that decays to zero by $t\!\approx\!5\times10^3$
(Fig.~\ref{fig:piandphi}e). A subtle but robust \emph{inverted temperature
ordering} appears in the transient: the $T\!=\!0.05$ curve has the slowest
$\pi$ approach and the \emph{highest} early-time $\phi$ peak, while
$T\!=\!2.0$ converges fastest. At low~$T$, the listener is slightly
more conservative and makes errors in both directions; higher~$T$ increases
overall permissiveness, paradoxically improving consistency. Since $\phi$
is already small, this cost is negligible and the net effect is beneficial.

\texttt{phi3:14b} \emph{(conservative listener).} $\pi(t)$ settles at
dramatically lower plateaux: ${\approx}\,0.75$ at $T\!=\!0.05$, falling
to ${\approx}\,0.30$ at $T\!=\!2.0$ (Fig.~\ref{fig:piandphi}c). At
$T\!=\!2.0$, 70\% of in-inventory interactions fail to trigger the collapse
they should. Meanwhile $\phi(t)$ is near zero at all temperatures
(Fig.~\ref{fig:piandphi}f), with only a faint persistent level
${\approx}\,0.03$--$0.05$ at $T\!=\!2.0$.
This places phi3 in the \emph{lazy-consolidation} limit
($\pi<1$, $\phi\approx 0$): the disordering channel is shut off, but the
ordering channel is heavily attenuated.

\begin{figure*}[t]
  \centering
  \includegraphics[width=0.325\textwidth]{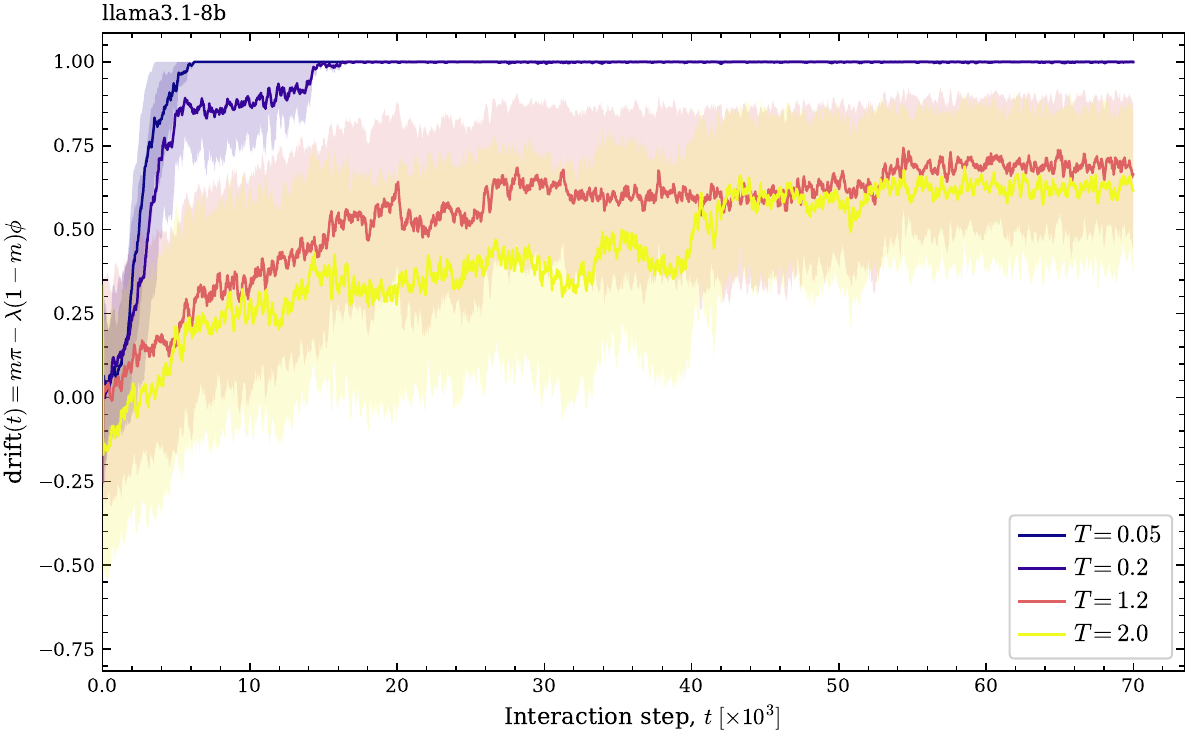}%
  \hfill
  \includegraphics[width=0.325\textwidth]{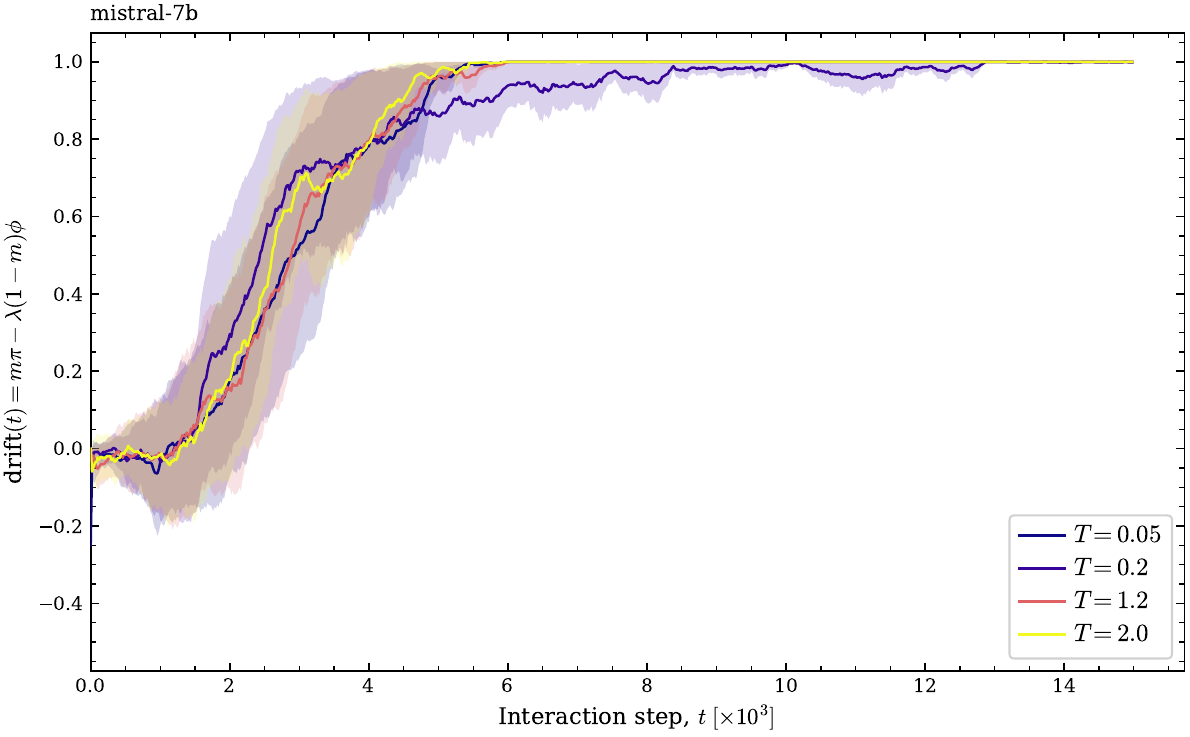}%
  \hfill
  \includegraphics[width=0.325\textwidth]{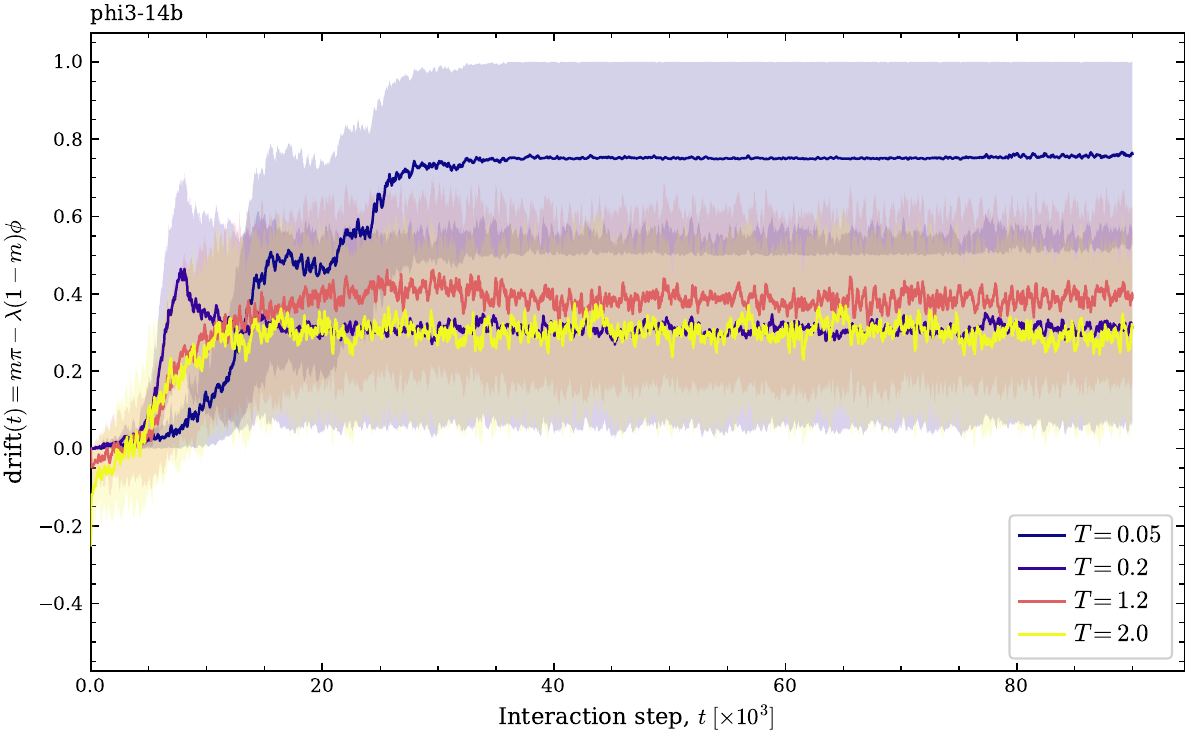}
  \caption{\label{fig:drift}%
  Drift proxy $\Delta(t) = m\pi - \lambda(1{-}m)\phi$ ($\lambda\!=\!1$)
  for (a)~\texttt{llama3.1:8b}, (b)~\texttt{mistral:7b},
  (c)~\texttt{phi3:14b}. Positive drift implies net ordering.}
\end{figure*}

\begin{figure*}[t]
  \centering
  \includegraphics[width=0.325\textwidth]{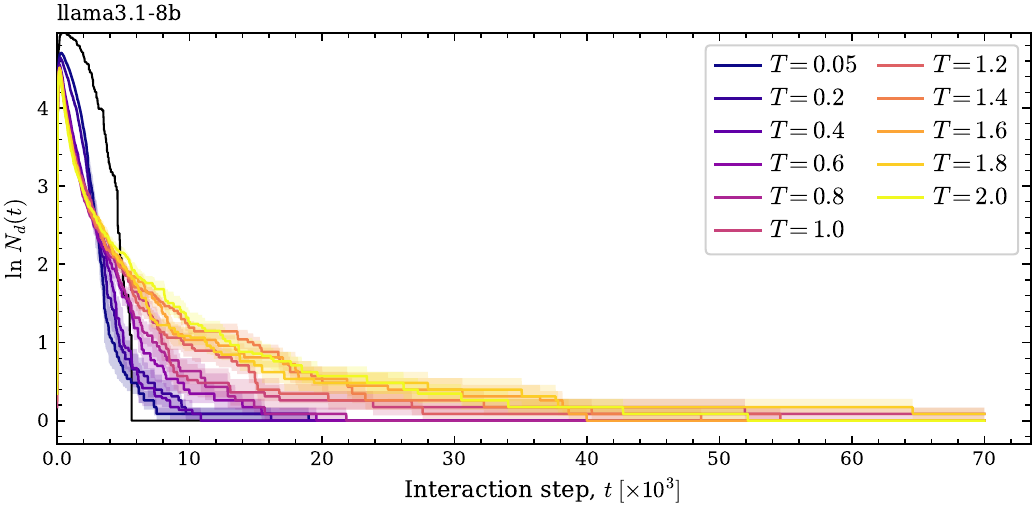}%
  \hfill
  \includegraphics[width=0.325\textwidth]{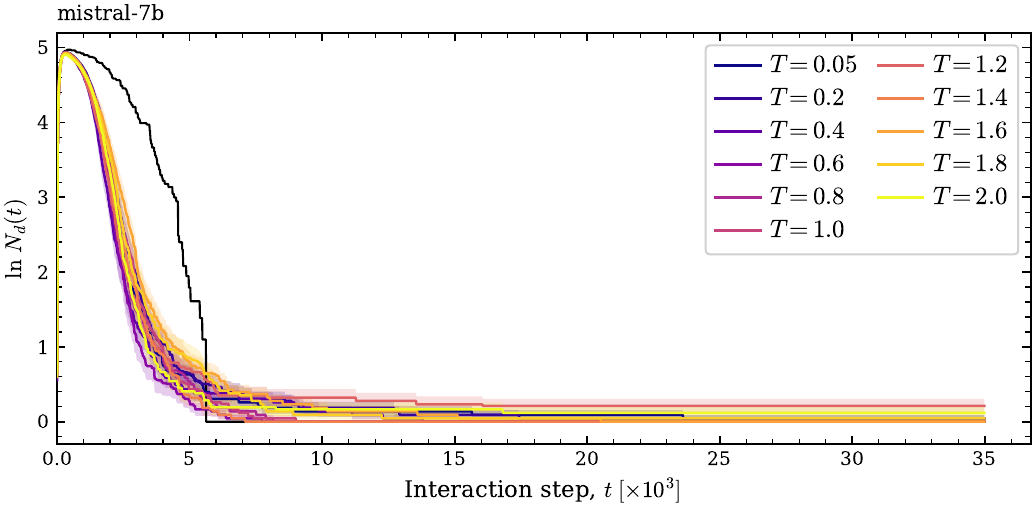}%
  \hfill
  \includegraphics[width=0.325\textwidth]{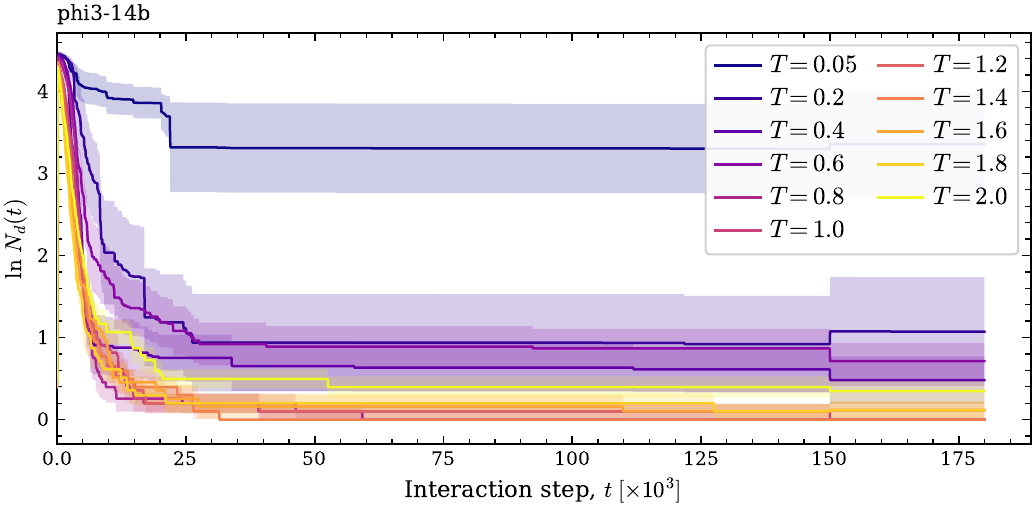}
  \caption{\label{fig:Nd}%
  Logarithm of the number of distinct words $\ln N_d(t)$ for $N\!=\!150$
  agents at all explored temperatures (colour-coded from dark$\,=\,$low~$T$
  to light$\,=\,$high~$T$). Solid black: \emph{deterministic} NG baseline.
  (a)~\texttt{llama3.1:8b}: standard ordering (high $T$ = slow).
  (b)~\texttt{mistral:7b}: near-deterministic, all $T$ bunched.
  (c)~\texttt{phi3:14b}: \emph{inverted} ordering (low $T$ = slowest).}
\end{figure*}

\begin{figure*}[t]
  \centering
  \includegraphics[width=0.325\textwidth]{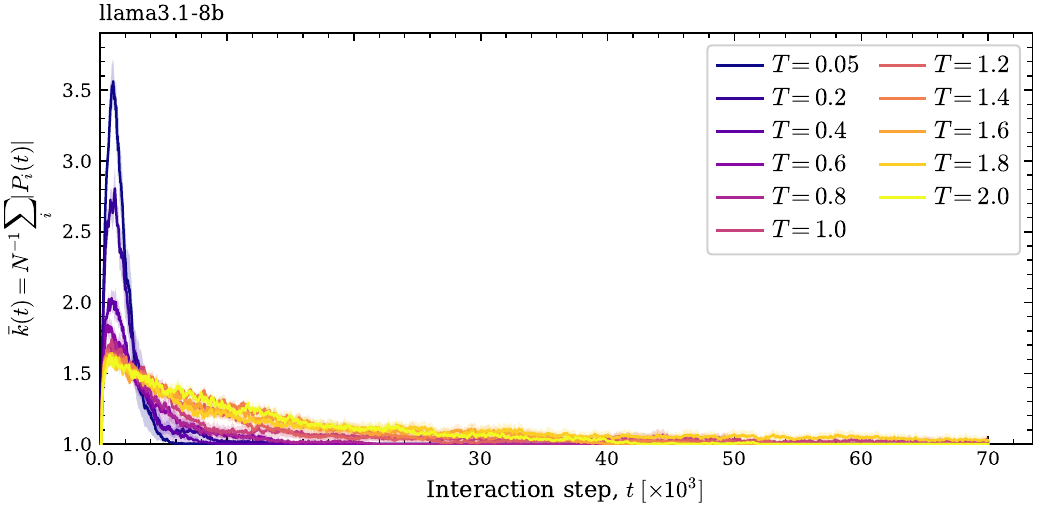}%
  \hfill
  \includegraphics[width=0.325\textwidth]{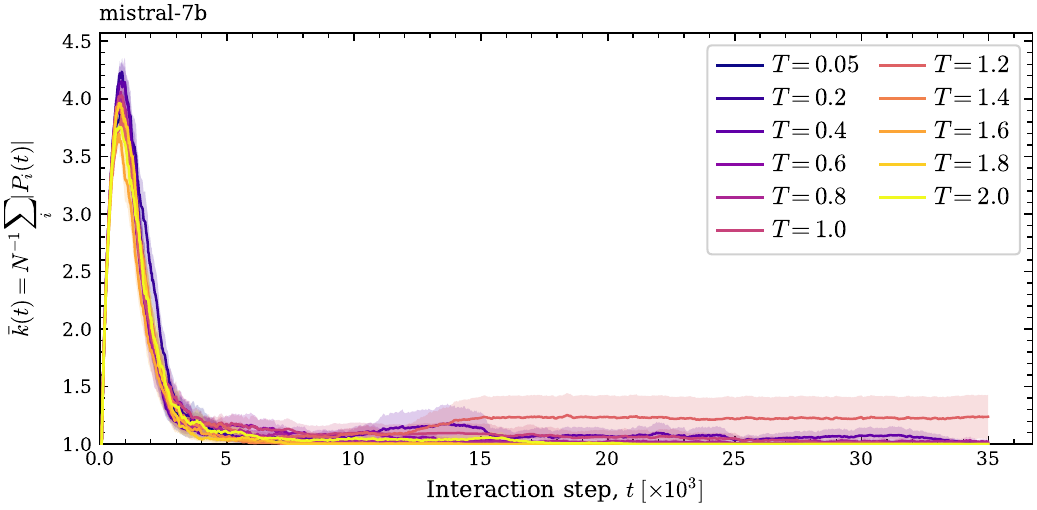}%
  \hfill
  \includegraphics[width=0.325\textwidth]{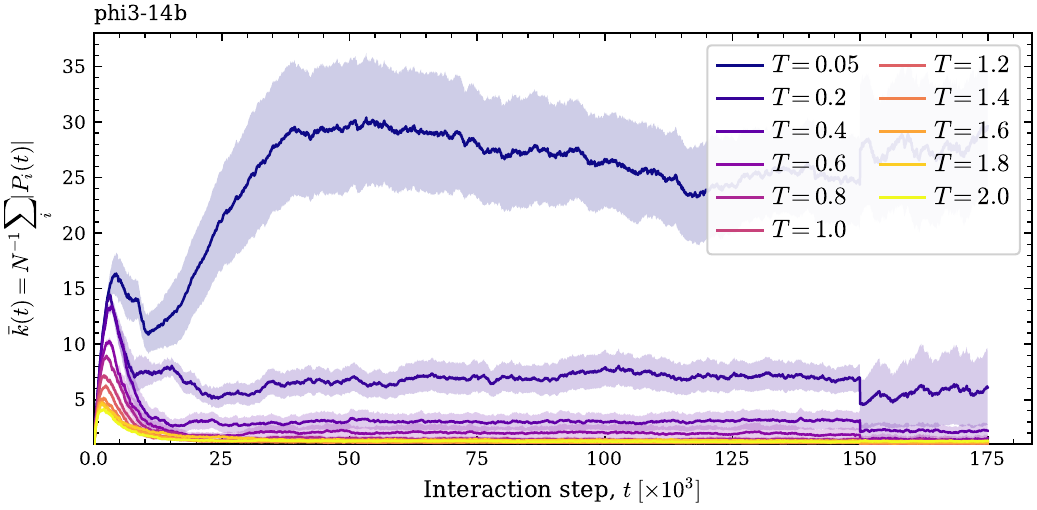}
  \caption{\label{fig:kbar}%
  Average inventory size $\bar{k}(t) = N^{-1}\sum_i |P_i(t)|$ for
  $N\!=\!150$ agents at all explored temperatures for
  (a)~\texttt{llama3.1:8b}, (b)~\texttt{mistral:7b},
  (c)~\texttt{phi3:14b}. Central lines are seed averages; shaded bands
  are one standard error. Note the different vertical scales: the phi3
  panel spans roughly an order of magnitude more than the other two.}
\end{figure*}

\begin{figure*}[t]
  \centering
  \includegraphics[width=0.325\textwidth]{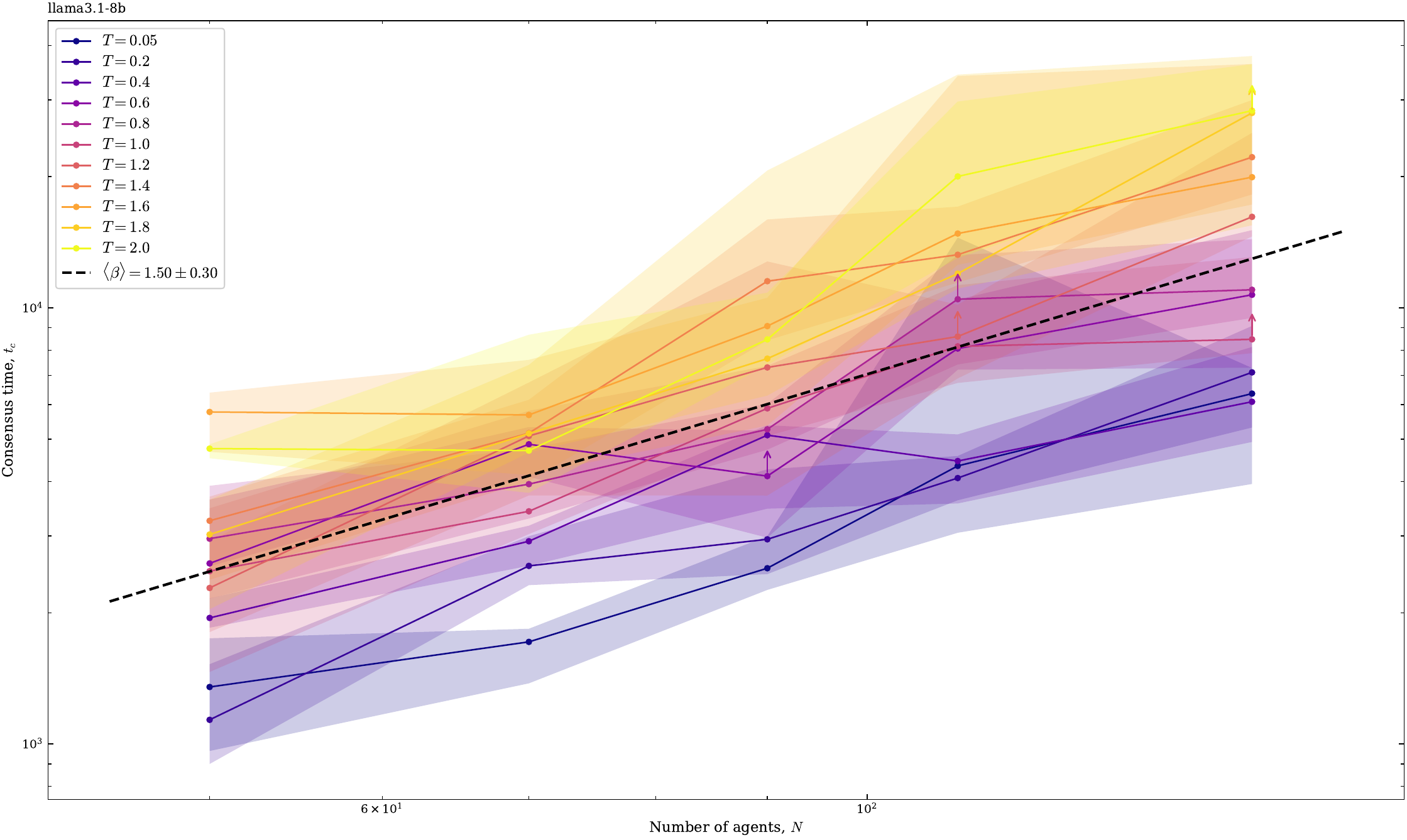}%
  \hfill
  \includegraphics[width=0.325\textwidth]{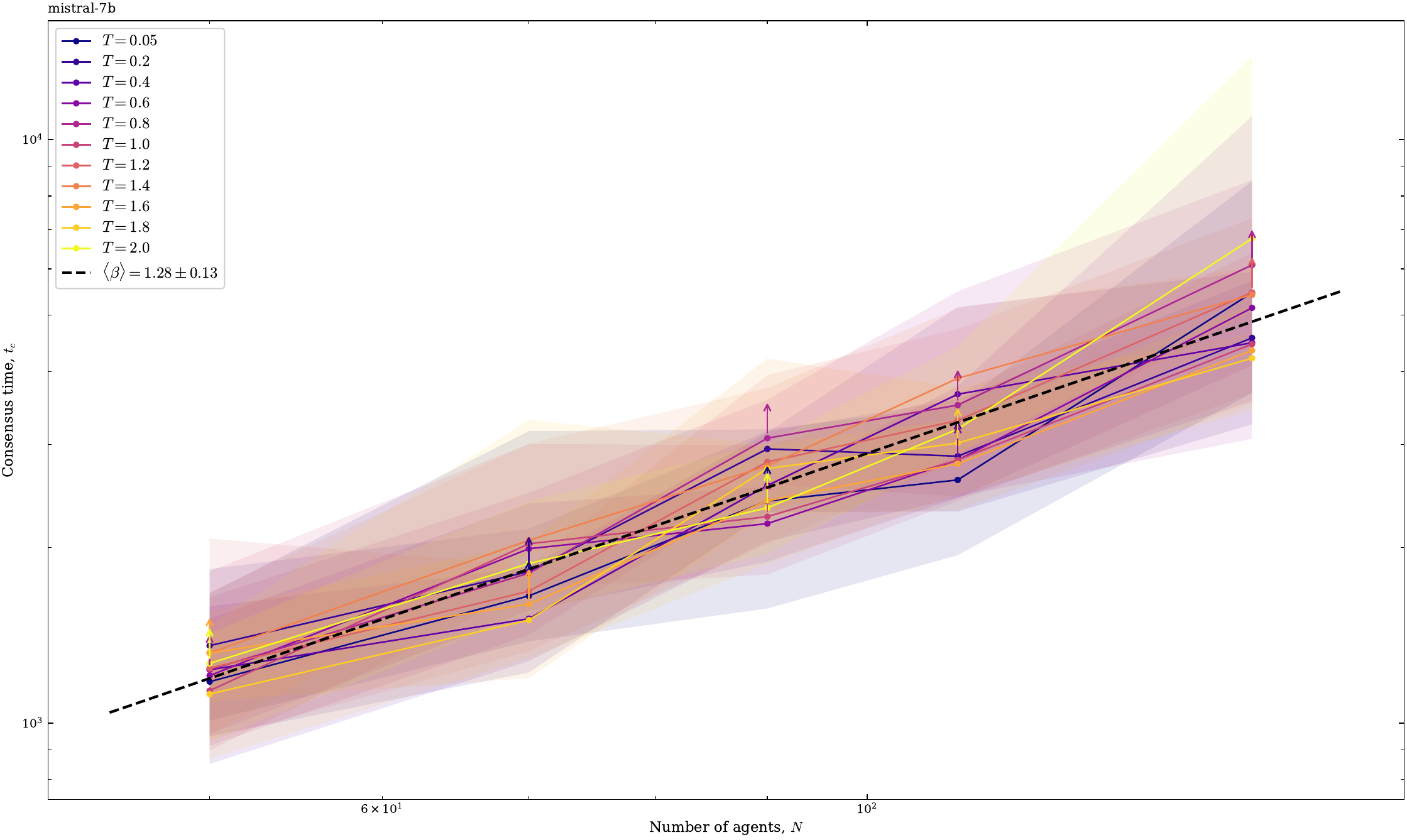}%
  \hfill
  \includegraphics[width=0.325\textwidth]{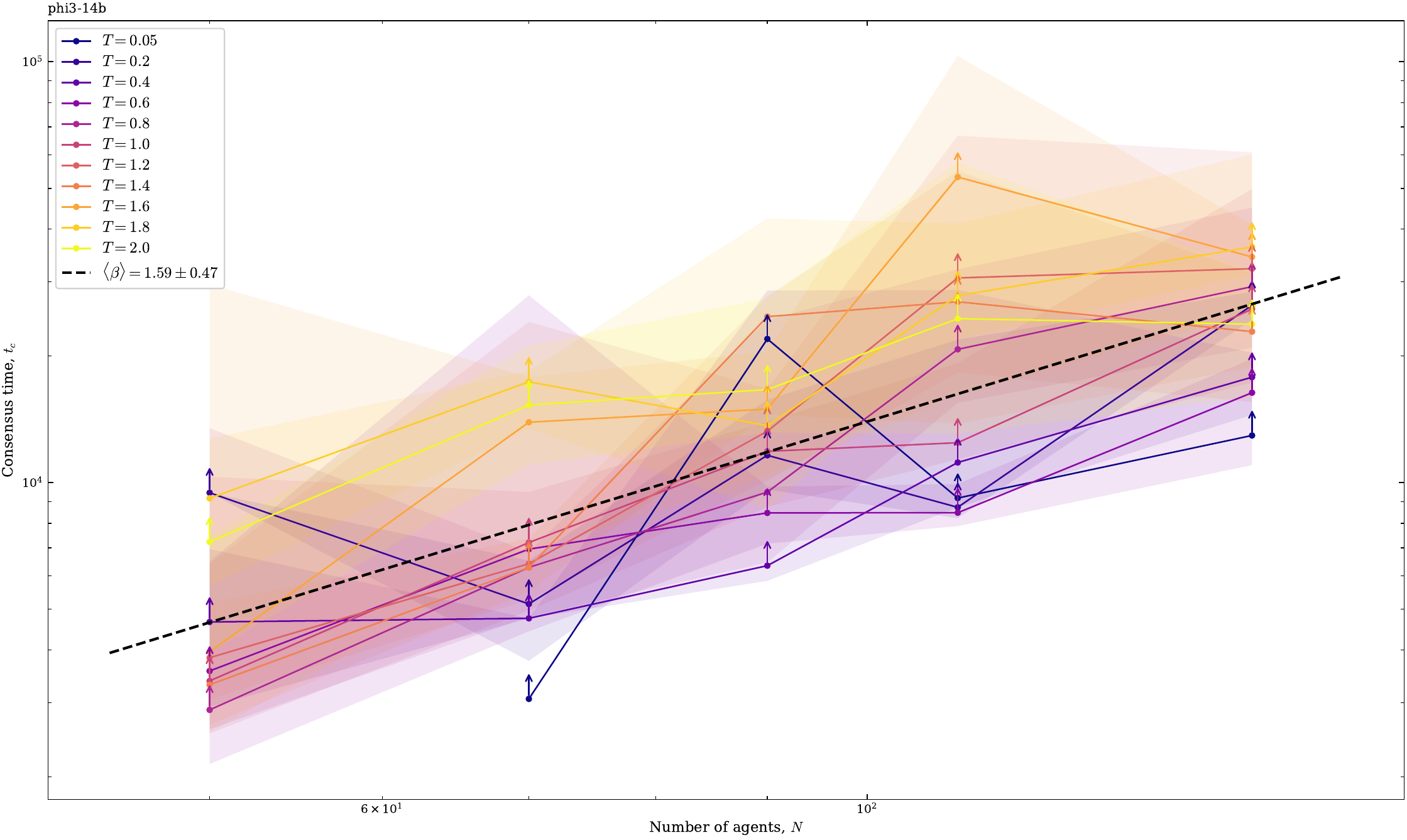}
  \caption{\label{fig:scaling}%
  Finite-size scaling of the consensus time $t_c$ vs.\ $N$ (log-log) for
  (a)~\texttt{llama3.1:8b},
  (b)~\texttt{mistral:7b},
  (c)~\texttt{phi3:14b}.
  Each colour is a different temperature. Central lines are medians
  across seeds, shaded bands are the interquartile range.
  Dashed black: average power-law guide $N^{\langle\beta\rangle}$.
  Upward arrows indicate lower bounds: at least one seed did not reach
  consensus within the simulation horizon, so the true $t_c$ is at least
  as large as the plotted value.}
\end{figure*}

\subsection{Drift and macroscopic convergence}
\label{sec:drift}

The drift proxy $\Delta(t)$, Eq.~\eqref{eq:drift}, translates the
microscopic rates into a net ordering tendency (Fig.~\ref{fig:drift}).

For \texttt{llama} (Fig.~\ref{fig:drift}a), $\Delta$ is strongly
temperature-dependent: at $T\!=\!0.05$ it reaches unity
(indistinguishable from \emph{deterministic} NG), while at $T\!=\!2.0$ it starts
weakly negative (repaint noise overwhelms consolidation) and climbs only
slowly to ${\approx}\,0.55$. The time spent at low or negative drift
maps directly onto the long plateaux in $N_d(t)$
(Fig.~\ref{fig:Nd}a).

For \texttt{mistral} (Fig.~\ref{fig:drift}b), all temperatures converge
to $\Delta\!=\!1$ within ${\sim}\,10^4$ steps. Surprisingly, $T\!=\!0.05$
shows the deepest early negative dip (${\approx}\,{-}0.25$), reflecting
the inverted-transient physics: the conservative low-$T$ listener
produces both missed collapses (lower $\pi$) and a slightly elevated
$\phi$ in the early disordered phase.

For \texttt{phi3} (Fig.~\ref{fig:drift}c), $\Delta$ is always positive
(no repaint noise) but never exceeds ${\approx}\,0.75$. The drift mirrors
$\pi$: weak but directed ordering without disordering competition.

The macroscopic convergence (Fig.~\ref{fig:Nd}) reveals three qualitatively
different responses of $N_d(t)$ to temperature:

\texttt{llama} (Fig.~\ref{fig:Nd}a) shows \emph{standard} temperature
ordering: higher~$T$ produces longer plateaux and slower convergence,
reflecting the monotonic growth of repaint noise $\phi(T)$.

\texttt{mistral} (Fig.~\ref{fig:Nd}b) shows all temperature curves
bunched tightly together and decaying \emph{faster} than the deterministic
baseline during the initial phase. The mild late-time splitting is
consistent with the inverted transient: low-$T$ curves have slightly longer
tails.

\texttt{phi3} (Fig.~\ref{fig:Nd}c) shows a \emph{strongly inverted}
temperature ordering: the lowest temperature ($T\!=\!0.05$, highest $\pi$)
is the \emph{slowest} to converge, with $N_d$ plateauing at
${\approx}\,7$ distinct words through $9\times 10^4$ steps without
reaching consensus. Higher temperatures converge faster despite having
lower~$\pi$. The mechanism is directly visible in the average inventory size
$\bar{k}(t) = N^{-1}\sum_i |P_i(t)|$ (Fig.~\ref{fig:kbar}). For phi3
at $T\!=\!0.05$, agents accumulate on average up to
${\approx}\,34$ words each around $t\!\approx\!7\times 10^4$ and
remain at a large-inventory plateau $\bar{k}\!\gtrsim\!30$
throughout the entire $1.75\times 10^5$-step simulation window
without reaching consensus; at $T\!=\!2.0$, by contrast,
$\bar{k}$ peaks briefly below ${\approx}\,4$ and decays to
$\bar{k}\!\approx\!1$ within ${\sim}\,10^4$ steps. llama and
mistral peak at $\bar{k}\!\lesssim\!5$ at every temperature and
collapse to $\bar{k}\!=\!1$ within $\mathcal{O}(10^4)$ steps.
This order-of-magnitude architecture-dependent difference in
per-agent inventory size is the microscopic origin of phi3's
inverted temperature ordering: the narrow $\pi\!\approx\!0.75$
channel must consolidate a much larger accumulated vocabulary at
low $T$ than at high $T$. A faint persistent
$\phi\!\approx\!0.04$ at $T\!=\!2.0$ may further help break
late-time deadlocks.

This inverted ordering demonstrates that the macroscopic convergence speed
is not determined by $\pi$ alone, but by the interplay of $\pi$ with the
inventory structure, a genuinely new feature of the LLM-NG that
emerges only when temperature is varied systematically.

\section{Finite-size scaling}
\label{sec:scaling}

In the \emph{deterministic} NG on a fully connected graph, $t_{\rm conv}\sim
N^{3/2}$~\cite{baronchelli2006sharp}. This scaling results from the interplay between the initial phase of word accumulation in the inventories of individual agents, the increase in correlations among the inventories of different agents, and the final coarsening collapse. We measure
$t_{\rm conv}(N,T)$ for $N\in\{50,70,90,110,150\}$ and fit
$t_{\rm conv}\sim N^{\beta(T)}$ (Fig.~\ref{fig:scaling}). The explored
range spans only about half a decade in~$N$, so the fitted exponents
should be interpreted as \emph{effective} quantities over this window
rather than as asymptotic critical exponents; a definitive determination
of the true asymptotic scaling would require sizes at least an order of
magnitude larger, beyond the reach of the present LLM-inference budget.

For \texttt{llama} (Fig.~\ref{fig:scaling}a), the effective exponent
varies from $\beta\!\approx\!1.3$ at low~$T$ to $\beta\!\approx\!2.0$
at $T\!=\!2.0$, with a temperature-averaged value
$\langle\beta\rangle\!=\!1.61\pm 0.28$. Over the accessible size range,
$\beta(T)$ reaches values above the canonical $3/2$ at high temperatures,
consistent with persistent repaint noise slowing the ordering drift.
Whether this reflects a genuine change of universality class or a slow
crossover to the deterministic $3/2$ asymptote at larger~$N$ cannot be
decided from the current data.

For \texttt{mistral} (Fig.~\ref{fig:scaling}b) we obtain
$\langle\beta\rangle\!=\!1.28\pm 0.13$, broadly consistent with an exponent
near the canonical $3/2$ and tighter than for the other two architectures.
This is the expected behaviour for a near-deterministic listener whose
microscopic rates barely depend on temperature.

For \texttt{phi3} (Fig.~\ref{fig:scaling}c) we find
$\langle\beta\rangle\!=\!1.59\pm 0.47$, consistent with $3/2$ within the
sizable uncertainty. The wide IQR bands and the persistent lower-bound
arrows in Fig.~\ref{fig:scaling}c are not statistical undersampling: with
$15$ seeds at horizon $1.75\times 10^{5}$ steps they reflect the
\emph{intrinsic} path-dependence of the conservative listener regime,
where low-$\pi$ dynamics make consensus trajectories strongly history
dependent. The reported value should be interpreted as a lower bound on
the true exponent, since seeds that fail to reach strict consensus within
the simulation horizon would, on average, raise it. It is worth noting
that phi3 is the empirical LLM realization of the lazy-consolidation
model of Ref.~\cite{baronchelli2007nonequilibrium}: to a very good
approximation ($\phi\!\approx\!0$ at every~$T$), phi3's dynamics
coincides with theirs under the identification
$\beta\!\leftrightarrow\!\pi(T)$. Their analytical prediction of a
consensus--fragmentation transition at $\beta_c\!=\!1/3$ is therefore a
direct, testable expectation for the phi3 regime (the same threshold emerges from the two-word mean field of Sec.~\ref{sec:meanfield} as the $\phi\!=\!0$ limit of the critical line): since phi3's measured
$\pi$ ranges from ${\approx}\,0.75$ at $T\!=\!0.05$ to ${\approx}\,0.30$
at $T\!=\!2.0$ (Sec.~\ref{sec:rates}), the threshold is nominally crossed
inside the accessible temperature window. We nevertheless observe strict
consensus at every temperature (Sec.~\ref{sec:discussion}), so testing the
prediction properly would require system sizes larger than those explored
here.

\section{Temperature response of the consensus time}
\label{sec:Tresponse}

\begin{figure*}[t]
  \centering
  \includegraphics[width=0.325\textwidth]{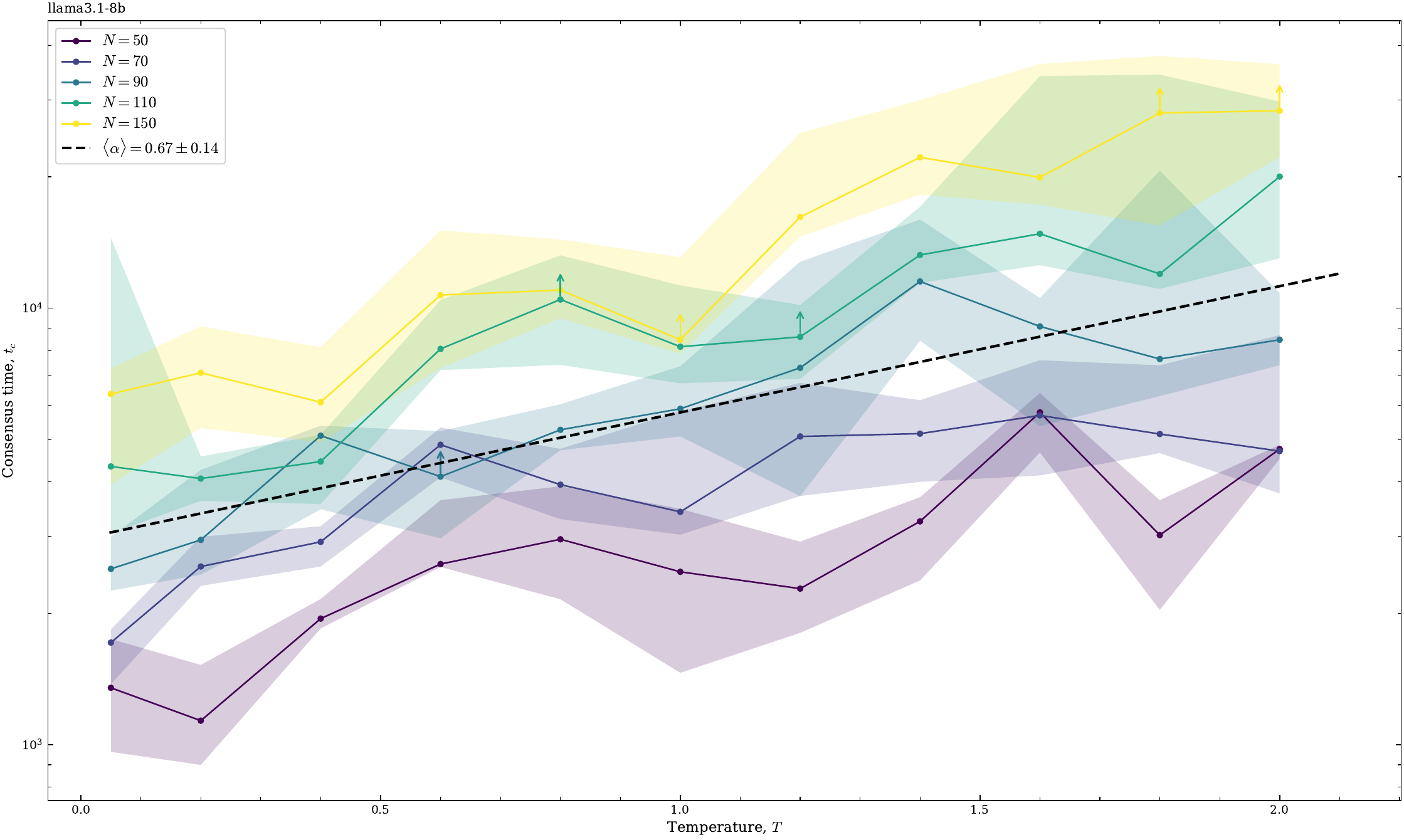}%
  \hfill
  \includegraphics[width=0.325\textwidth]{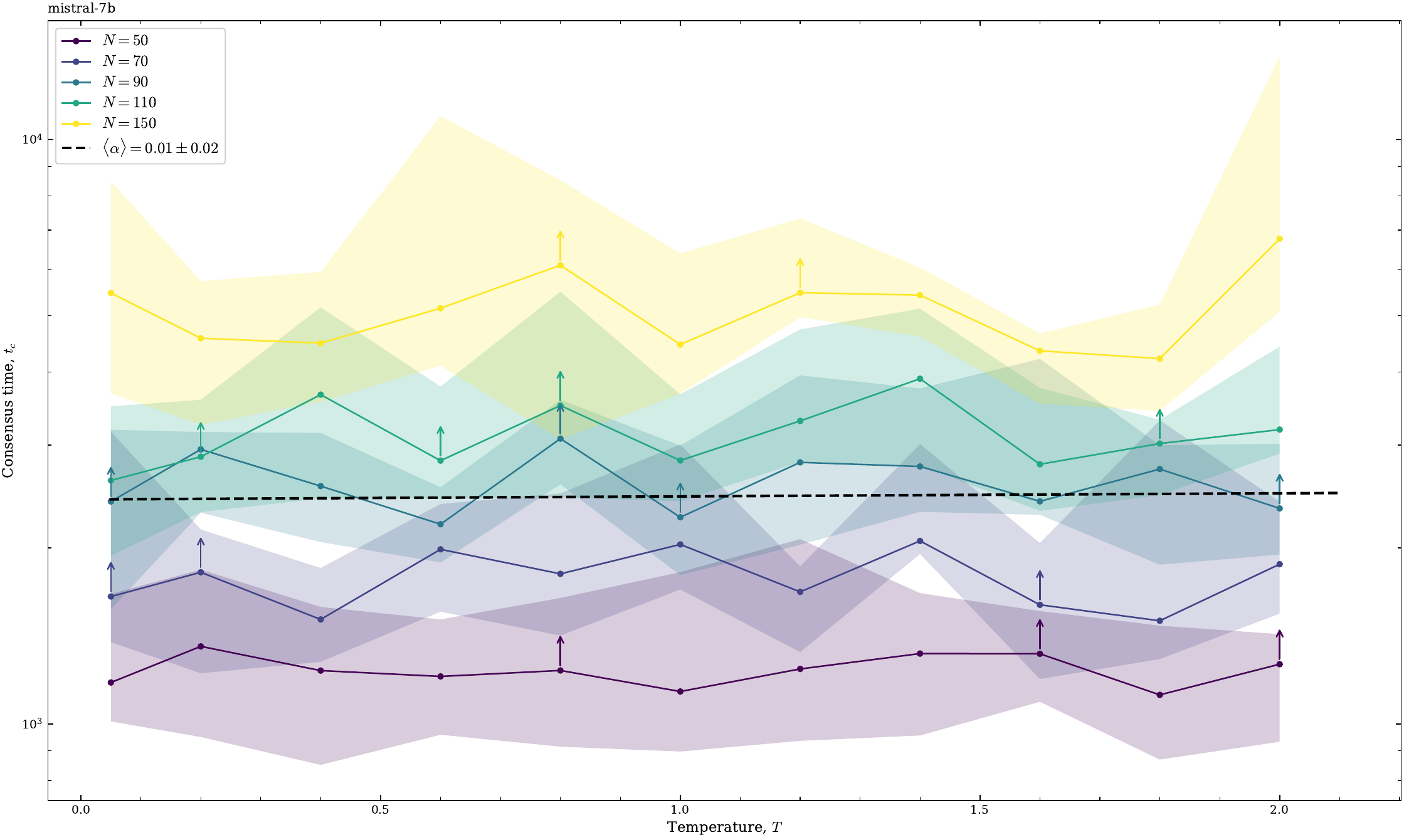}%
  \hfill
  \includegraphics[width=0.325\textwidth]{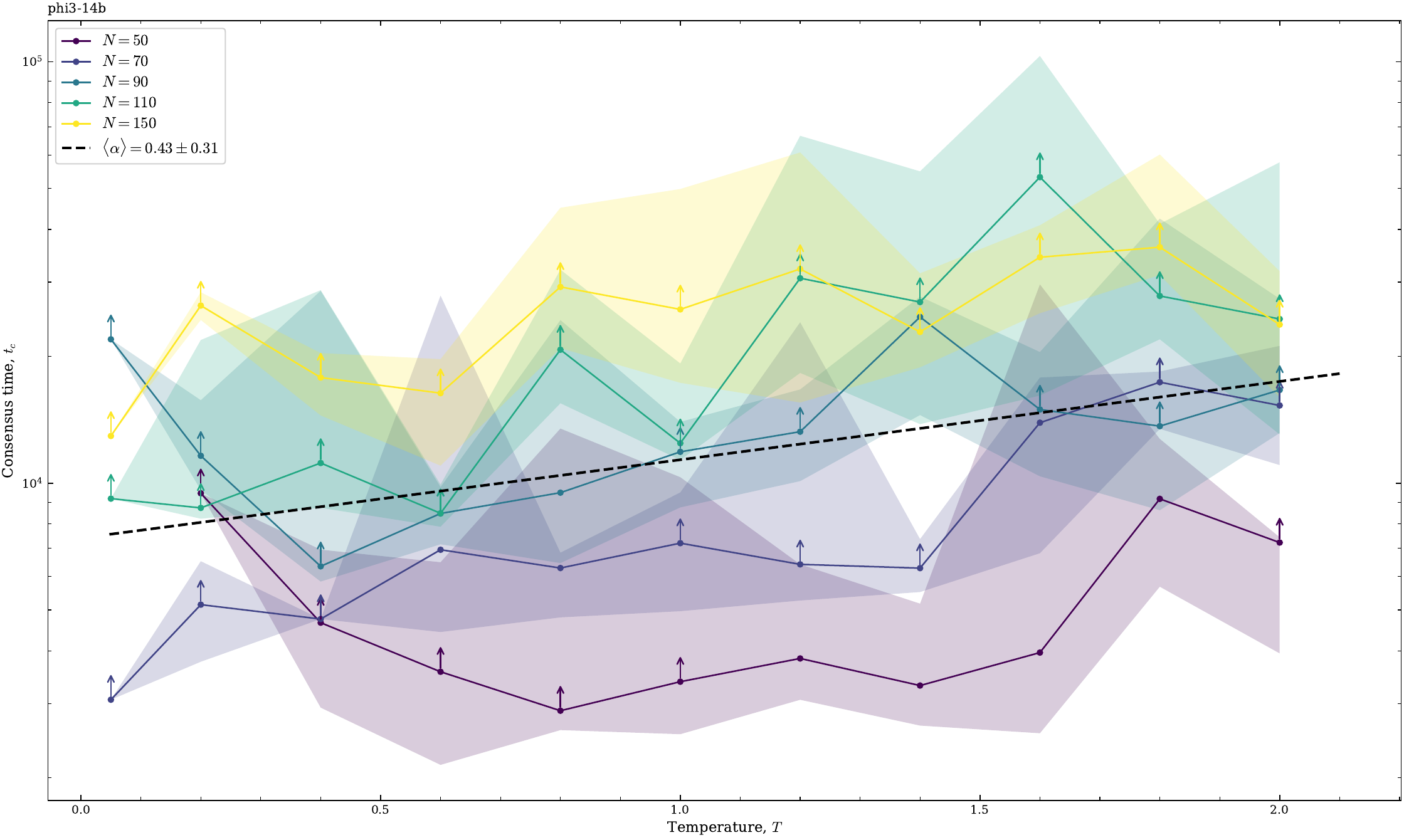}
  \caption{\label{fig:tcvsT}%
  Temperature response of the consensus time $t_c$ vs.\ $T$ (log-y, linear-x)
  for (a)~\texttt{llama3.1:8b}, (b)~\texttt{mistral:7b},
  (c)~\texttt{phi3:14b}. Each colour is a different agent count
  $N\!\in\!\{50,70,90,110,150\}$; central lines are medians across seeds and
  shaded bands span the interquartile range. Dashed black: average
  exponential guide $t_c\!\sim\!e^{\langle\alpha\rangle\,T}$. The fitted
  rates are
  $\langle\alpha\rangle\!=\!0.67\pm 0.14$ for \texttt{llama},
  $\langle\alpha\rangle\!=\!0.01\pm 0.02$ for \texttt{mistral}, and
  $\langle\alpha\rangle\!=\!0.43\pm 0.31$ for \texttt{phi3}.
  Upward arrows indicate lower bounds: at least one seed did not reach
  consensus within the simulation horizon.}
\end{figure*}

While Fig.~\ref{fig:scaling} examines how $t_c$ scales with $N$ at fixed
temperature, the complementary question is how $t_c$ depends on $T$ at
fixed~$N$. We address it in Fig.~\ref{fig:tcvsT}, which displays
$t_c(T)$ for $N\in\{50,70,90,110,150\}$ across the three architectures.
Because $t_c$ varies multiplicatively with~$T$, we fit each curve in
log-linear form,
\begin{equation}
  \ln t_c(T) \,=\, \ln A \,+\, \alpha\, T\,,
  \label{eq:alpha}
\end{equation}
so that $t_c(T)\!=\!A\,e^{\alpha T}$ and the slope~$\alpha$ has units of
inverse temperature. The sign and magnitude of~$\alpha$ provide a direct,
single-number summary of the temperature sensitivity of the macroscopic
consensus dynamics: $\alpha\!>\!0$ means $t_c$ grows with~$T$,
$\alpha\!\approx\!0$ means $T$-independence, and $\alpha\!<\!0$ would
indicate accelerated convergence with~$T$. The exponential form mirrors
the power-law guide of Fig.~\ref{fig:scaling} (a straight line on log-y)
and produces a clean dashed-line overlay on each panel of
Fig.~\ref{fig:tcvsT}.

The three measured rates differ qualitatively across architectures and map
cleanly onto the $(\pi,\phi)$ regimes identified in
Sec.~\ref{sec:rates}.

\paragraph*{llama (permissive listener, $\alpha\!=\!0.67\pm 0.14$).}
The exponent is clearly positive and statistically significant: across the
explored range $t_c$ grows by a factor $e^{0.67\times 2.0}\!\approx\!4$.
This is the macroscopic fingerprint of the repaint-dominated regime: as~$T$
rises, $\phi(T)$ grows and the disordering channel becomes increasingly
active; the consensus time inherits this growth at an exponential rate. The
exponential, rather than linear, dependence on~$T$ is itself informative; it
suggests that $t_c$ is sensitive to the \emph{cumulative} effect of many
repaint events along an ordering trajectory, rather than to any single
rate-limiting step.

\paragraph*{mistral (near-deterministic listener, $\alpha\!=\!0.01\pm 0.02$).}
The exponent is indistinguishable from zero: $t_c$ is statistically
\emph{independent} of decoding temperature across a $40\times$ range in~$T$.
This is the macroscopic fingerprint of the near-deterministic regime:
with $\pi\!\approx\!1$ and $\phi\!\approx\!0$ at every~$T$
(Sec.~\ref{sec:rates}), neither microscopic rate has room to move, and the
resulting consensus dynamics is effectively decoupled from the LLM's primary
stochasticity knob. A similar architecture-dependent insensitivity to
decoding temperature has been reported in a related statistical-physics
study of LLM populations on
lattices~\cite{denobili2026collectivealignmentllmmultiagent},
suggesting that this ``temperature blindness'' is a robust feature of
certain LLM architectures across distinct collective-dynamics settings.

\paragraph*{phi3 (conservative listener, $\alpha\!=\!0.43\pm 0.31$).}
The central value is positive and moderate, but the uncertainty band
overlaps zero. Together with the wide IQR shading in
Fig.~\ref{fig:tcvsT}c, this reflects the genuine path-dependence of the
conservative regime: low-$\pi$ dynamics is intrinsically variable from
seed to seed, and the temperature response is more subtle than for the
other two architectures. The trend is mediated by the
inventory-diversity mechanism described in Sec.~\ref{sec:drift} rather
than by a direct change in the microscopic rates, since~$\phi$ remains
near zero at every~$T$.

The pair $(\beta,\alpha)$ thus provides a compact two-dimensional
fingerprint of each architecture in the multi-agent NG: $\beta$ measures how
the consensus time scales with system size, $\alpha$ how it responds to
decoding temperature. Their joint values cleanly separate the three regimes
(Table~\ref{tab:regimes}).

\section{Mean-field theory of the two-rate dynamics}
\label{sec:meanfield}

At fixed decoding temperature, the LLM listener can be approximated by
the effective two-rate rule
\begin{equation}
  q(\text{YES}\mid P_j, w) \;=\;
  \begin{cases}
    \pi, & w\in P_j,\\[2pt]
    \phi, & w\notin P_j,
  \end{cases}
  \label{eq:tworate}
\end{equation}
with $\pi$ and $\phi$ treated as constants, in practice the plateau
values $\bar\pi(T)$ and $\bar\phi(T)$ of Sec.~\ref{sec:rates}. This
defines a stochastic NG with two acceptance channels that contains the
deterministic NG ($\pi\!=\!1$, $\phi\!=\!0$) and the stochastic
negotiation model of Ref.~\cite{baronchelli2007nonequilibrium}
($\pi\!=\!\beta$, $\phi\!=\!0$) as special cases, and for which exact complete-graph mean-field equations can be written for the densities of agents carrying each possible inventory. The resulting
$(2^m{-}1)$-dimensional hierarchy is not solvable in closed form, but it admits a natural approximate closure for the inventory-size
distribution; both are given in Appendix~\ref{app:meanfield}. The
critical line, however, follows analytically from the two-word sector, to which we now turn.

The ordering mechanism is exposed by the two-word sector. When only two
words $A$ and $B$ compete, each agent is in one of three states, $A$,
$B$, or $AB$, with fractions $x$, $y$, and $z=1-x-y$. On the complete
graph the mean-field equations read
\begin{align}
  \dot{x} &= -(1-\phi)\,xy + \tfrac{3\pi-1}{2}\,xz + \phi\,yz + \pi z^2,
  \label{eq:mfx}\\
  \dot{y} &= -(1-\phi)\,xy + \tfrac{3\pi-1}{2}\,yz + \phi\,xz + \pi z^2,
  \label{eq:mfy}
\end{align}
reducing to the standard NG mean field for $\pi\!=\!1$, $\phi\!=\!0$.
Introducing the magnetization-like order parameter $u = x - y$ and
subtracting Eq.~\eqref{eq:mfy} from Eq.~\eqref{eq:mfx} gives
\begin{equation}
  \dot{u} \;=\; \frac{z}{2}\,\bigl(3\pi - 2\phi - 1\bigr)\,u\,.
  \label{eq:ordering}
\end{equation}
The symmetric state is unstable, and one word is amplified over the
other, when
\begin{equation}
  R \;\equiv\; 3\pi - 2\phi - 1 \;>\; 0,
  \quad\text{i.e.}\quad
  \pi \;>\; \pi_c(\phi) = \frac{1+2\phi}{3}\,.
  \label{eq:Rcrit}
\end{equation}
For $\phi\!=\!0$ this recovers the known threshold $\beta_c\!=\!1/3$ of
the stochastic NG~\cite{baronchelli2007nonequilibrium}; the
false-positive channel shifts the threshold upward along the critical
line $\pi_c(\phi)$, quantifying how repaint noise obstructs ordering.
The quantity $R$ thus plays the role of the effective distance from the
transition, replacing the scalar $\beta-\beta_c$, and provides an
analytical counterpart to the phenomenological drift proxy of
Eq.~\eqref{eq:drift}.

Evaluated on the measured plateau rates, $R$ rationalizes the three
regimes. For \texttt{mistral}, $\bar\pi\!\approx\!1$ and
$\bar\phi\!\approx\!0$ give $R\!\approx\!2$, the maximum possible
value: the dynamics sits deep inside the ordering region at every
temperature, which is precisely the temperature blindness of
Sec.~\ref{sec:Tresponse}. For \texttt{llama}, increasing $T$ lowers
$\pi$ and raises $\phi$; both changes decrease $R$, predicting the
observed monotonic slowdown, and the elevated early-time $\phi$ at
$T\!=\!2.0$ transiently drives $R$ negative, consistent with the
negative drift transient of Fig.~\ref{fig:drift}a. For \texttt{phi3},
$\bar\phi\!\approx\!0$ reduces the condition to $\pi>1/3$: the measured
plateau falls from $\bar\pi\!\approx\!0.75$ ($R\!\approx\!1.25$) at
$T\!=\!0.05$ to $\bar\pi\!\approx\!0.30$ ($R\!\approx\!-0.1$) at
$T\!=\!2.0$, nominally crossing the critical line inside the explored
temperature window. That strict consensus is nevertheless reached at
all temperatures (Sec.~\ref{sec:discussion}) is consistent with the
mean-field transition being sharp only as $N\to\infty$ and with the
time dependence of the measured rates, and identifies phi3 at high~$T$
as the natural setting in which to search for a fragmented phase at
larger system sizes. By construction, the two-word reduction does not capture the inventory-size effects that dominate phi3 at low~$T$ (Sec.~\ref{sec:drift}); the inventory-size closure of
Appendix~\ref{app:meanfield} is the natural starting point to include them.

\section{Discussion and outlook}
\label{sec:discussion}

The central result of this work is that two conditional rates, $\pi(T)$
and $\phi(T)$, provide a compact microscopic parametrisation of the
LLM-NG that captures its leading-order macroscopic phenomenology and
cleanly organizes the observed architecture-dependent behaviour into
three qualitatively distinct regimes (Table~\ref{tab:regimes}). The
$(\pi,\phi)$ pair is not a full microscopic theory: it averages over
inventory size, over word identity, and over per-agent heterogeneity,
and phi3 already provides an example (Sec.~\ref{sec:drift}) in which
macroscopic convergence is co-determined by $\pi$ and by the inventory
structure it generates. Nevertheless, the sign of the drift proxy
$\Delta(t)$ built from $(\pi,\phi)$ tracks the qualitative
ordering / disordering balance of every architecture we tested, and
the pair suffices as a first-order diagnostic and classification tool.

\begin{table}[b]
\caption{\label{tab:regimes}%
Classification of the three LLM architectures by their $(\pi,\phi)$
response, effective scaling exponent $\langle\beta\rangle$ over
$N\!\in\![50,150]$, temperature-sensitivity $\langle\alpha\rangle$, and
dominant slowdown mechanism. Arrows $\searrow(T)$ and $\nearrow(T)$ denote
decrease and increase with temperature.}
\begin{ruledtabular}
\begin{tabular}{lccccl}
Model & $\pi$ & $\phi$ & $\langle\beta\rangle$ & $\langle\alpha\rangle$ & Slowdown \\
\colrule
\texttt{llama3.1:8b} & $\searrow(T)$ & $\nearrow(T)$ & $1.61$ & $0.67$ & repaints \\
\texttt{mistral:7b} & ${\approx}\,1$ & ${\approx}\,0$ & $1.28$ & $0.01$ & (weak) \\
\texttt{phi3:14b} & low, $\searrow(T)$ & ${\approx}\,0$ & $1.59$ & $0.43$ & missed collapses \\
\end{tabular}
\end{ruledtabular}
\end{table}

These three regimes are not imposed by hand but \emph{emerge} from the
interaction between the LLM architecture and the decoding temperature. The
$(\pi,\phi)$ decomposition acts as a bridge between the ``black box'' of LLM
inference and the well-understood physics of the NG.

Our work is complementary to but distinct from recent LLM-NG studies in
four specific ways. First, Ref.~\cite{ashery2025emergent} and
Ref.~\cite{flint2025groupsizeeffectscollective} characterise the
\emph{macroscopic} outcome (does consensus emerge, on which name, with
what bias) at a single fixed temperature $T\!=\!0.5$. We instead
decompose each microscopic interaction into the conditional rates
$(\pi,\phi)$ and trace how decoding temperature reshapes them. Second,
their analysis treats the LLM as an unresolved black box and infers
asymmetries from observed bias; we expose the underlying in-inventory and out-inventory channel structure that produces those asymmetries. Third, we
elevate decoding temperature from a fixed hyperparameter to a tunable
control parameter and quantify its effect through two complementary
scalar diagnostics, $\beta(T)$ and $\alpha(N)$. Fourth, we identify three
qualitatively distinct architecture-dependent regimes that produce
different consensus dynamics through different microscopic mechanisms.
Together, these moves convert LLM-NG from a phenomenological observation
that consensus emerges into a microscopically resolved
statistical-physics problem. Our results are also complementary to those
of Ref.~\cite{demarzo2025aiagentscoordinatehuman}, which established the
existence of architecture-dependent critical group sizes, and to those
of Ref.~\cite{demarzo2026conformitygeneratescollectivemisalignment},
which decomposes conformity into competing effective forces; in both
cases, the $(\pi,\phi)$ rates provide a temporally resolved diagnostic
that complements those purely macroscopic or aggregate-state
characterizations.

Two further studies complement our findings by varying different knobs of
the LLM-NG. Mehdizadeh and Hilbert~\cite{mehdizadeh2026exploring} study a
networked Naming Game among LLM agents and show that agent memory depth
interacts with network topology in a sign-flipping way: longer memory slows
convergence in decentralized networks but accelerates the fragmented
settling of centralized ones. Their work fixes population size and varies
topology and memory, providing a natural counterpart to our
temperature-based control parameter; whether their effects persist under
finite-size scaling, and how they interact with the $(\pi,\phi)$
decomposition, remains open. Separately, Zhang
\emph{et al.}~\cite{zhang2025sign} show that imposing lightweight schema
structure on the communication channel itself, rather than tuning a decoding
hyperparameter, can accelerate naming-game convergence by up to
$5.8\times$. Together with our results, this points to a broader picture in
which convention formation in LLM populations can be steered through several
largely independent levers, decoding temperature, memory depth, network
topology, and communication schema, each acting on a different part of the
underlying $(\pi,\phi)$ channel structure.

Several features are worth emphasising. First, temperature does not
universally slow consensus: for \texttt{mistral}, increasing~$T$ paradoxically
\emph{improves} consistency by suppressing the low-$T$ conservative
transient; for \texttt{phi3}, high~$T$ accelerates convergence despite
lowering~$\pi$, because it also reduces inventory diversity. The macroscopic
effect of decoding temperature is thus \emph{architecture-dependent},
mediated by the model-specific balance of the ordering and disordering
channels. This cautions against treating temperature as a universal ``noise
knob'' in multi-agent LLM systems.

Second, from a statistical-physics perspective, the relationship between
our $(\pi,\phi)$ framework and the stochastic negotiation model of Baronchelli
\emph{et al.}~\cite{baronchelli2007nonequilibrium} is more than an
analogy: their commitment probability $\beta$ is mathematically identical
to our $\pi$, and their model corresponds exactly to the $\phi\!=\!0$
slice of the $(\pi,\phi)$ plane. The essential conceptual difference is
in the \emph{origin} of the stochasticity. In their setup, $\beta$ is
an external, hand-tuned scalar, chosen by the modeller and swept across
$[0,1]$ to trace out a phase diagram; the update rule fixes
$\phi\!=\!0$ by construction. In ours, both $\pi(T)$ and $\phi(T)$ are
\emph{emergent}: they are properties of the LLM listener that we
measure a posteriori from multi-agent simulations, and their values,
their $T$-dependence, and even their qualitative shapes are set by the
architecture. This shift, from an external stochasticity parameter to
an emergent one, is what turns a mathematically clean toy model into a
diagnostic tool for real LLM populations. As a corollary, decoding
temperature moves both rates along architecture-dependent
trajectories in the two-dimensional $(\pi,\phi)$ plane, whereas the
phase transition of Ref.~\cite{baronchelli2007nonequilibrium} is
inherently a one-dimensional phenomenon along the $\phi\!=\!0$ edge,
of which phi3 is the empirical embodiment (Sec.~\ref{sec:scaling}).
Notably, no fragmented phase was observed for any model at the explored
temperatures: strict $1$-consensus is always reached on the fully connected
graph, even for phi3 at $T\!=\!2.0$, where $R\!\approx\!-0.1$ lies nominally
below the critical line (Sec.~\ref{sec:meanfield}). As discussed there,
this is consistent with a transition that sharpens only as $N\to\infty$, and with the group-size scenarios of Refs.~\cite{flint2025groupsizeeffectscollective,demarzo2025aiagentscoordinatehuman},
where fragmentation emerges only above model-dependent population thresholds, beyond the sizes explored here.

Third, over the accessible size range $N\!\in\![50,150]$ the effective
scaling exponent $\beta(T)$ varies with temperature, reaching values above
$3/2$ for \texttt{llama} at high~$T$. Discriminating a genuine change of
universality class from a slow crossover to the deterministic asymptote
would require substantially larger systems, a natural target for future
work.

More broadly, the $(\pi,\phi)$ framework is not specific to the naming
game; any binary decision made by an LLM agent in the presence of a
ground-truth state can be decomposed analogously. Recent
statistical-physics analyses of LLM populations on
lattices~\cite{denobili2026collectivealignmentllmmultiagent} and in
conformity-driven opinion
dynamics~\cite{demarzo2026conformitygeneratescollectivemisalignment} have
similarly relied on decompositions of the LLM response into competing
effective parameters (cooperative coupling vs.\ intrinsic bias; majority
force vs.\ individual preference). Asch-type conformity experiments on
multimodal LLM agents~\cite{bellina2026conformity}, where a binary
judgement with a known ground truth is flipped by social pressure,
provide a particularly close setting: the probability of yielding to
the group plays the same role as our false-positive rate $\phi$. Our
$(\pi,\phi)$ rates fit naturally into this emerging framework as a
temporally-resolved diagnostic.

These findings carry direct implications for decentralized and
self-organizing LLM populations. As multi-agent LLM systems are
increasingly deployed without central coordination, for example in
federated learning, autonomous negotiation, distributed scientific
discovery, and collective decision making, the question of whether and
how fast they converge on shared conventions becomes operational. Our
results provide three quantitative answers. First, the $(\pi,\phi)$
rates can be measured offline through a small set of probing prompts, so
that a system designer can determine which regime a given LLM occupies
before deploying it. Second, the rate $\alpha$ tells the designer whether
decoding temperature is a useful design knob: for \texttt{mistral}-like
architectures ($\alpha\!\approx\!0$) tuning $T$ has essentially no
effect on consensus speed, while for \texttt{llama}-like architectures
($\alpha\!\approx\!0.7$) $T$ changes $t_c$ by a factor of about four
across the accessible range. Third, the temperature-dependence of
$\beta(T)$ means that the effective collective dynamics of an LLM swarm
can be shaped through the joint choice of architecture and decoding
temperature, opening the door to physics-informed selection of LLMs for
specific multi-agent tasks. We view the identification of distinct
listener archetypes and their mapping onto macroscopic consensus dynamics
as a first step toward a statistical-physics taxonomy of LLM-agent
behaviour, of which the $(\pi,\phi,\beta,\alpha)$ quadruple is a candidate
compact ``datasheet'' summarising an architecture's collective properties,
much as critical exponents summarise a universality class.

Natural extensions include: heterogeneous temperatures (quenched
disorder), structured interaction topologies, multi-object naming, the
inclusion of committed
minorities~\cite{xie2011social,ashery2025emergent}, and
the exploration of model-native commitment signals (e.g.\ confidence
scores) as additional control parameters. The mean-field theory of Sec.~\ref{sec:meanfield} opens several
analytical directions: the analysis of the inventory-size hierarchy of Appendix~\ref{app:meanfield}, which would capture the large-inventory effects that dominate the phi3 regime at low~$T$; the structure of the possible fragmented states of the two-rate dynamics; the scaling of the convergence time with the distance $R$ from the critical line; and the finite-$N$ rounding of the transition, which the phi3 high-$T$ regime is best positioned to probe.

\appendix
\section{Mean-field equations for arbitrary vocabulary and
inventory-size closure}
\label{app:meanfield}

Let the active vocabulary contain $m$ words,
$\Omega_m=\{1,\dots,m\}$, and let $n_S(t)$ be the fraction of agents
with inventory $S\subseteq\Omega_m$, $S\neq\emptyset$, normalized as
$\sum_{S} n_S = 1$. On the complete graph, in the limit $N\to\infty$,
the two-rate rule of Eq.~\eqref{eq:tworate} induces the exact
mean-field dynamics
\begin{widetext}
\begin{equation}
  \dot{n}_S \;=\; \sum_{A,B\neq\emptyset} n_A n_B\, \frac{1}{|A|}
  \sum_{w\in A}
  \Bigl[\, q_B(w)\,\bigl(2\,\delta_{S,\{w\}} - \delta_{S,A} - \delta_{S,B}\bigr)
  \;+\; \bigl(1-q_B(w)\bigr)\bigl(\delta_{S,B\cup\{w\}} - \delta_{S,B}\bigr)
  \Bigr],
  \label{eq:master}
\end{equation}
\end{widetext}
where $A$ and $B$ are the speaker and listener inventories, the
speaker chooses $w\in A$ uniformly, $q_B(w)=\pi$ if $w\in B$ and
$q_B(w)=\phi$ otherwise, and $\delta_{S,X}$ is the Kronecker delta on inventories. The first bracket describes YES events, in which both agents collapse to $\{w\}$; the second describes NO events, in which the speaker is unchanged and the listener learns $w$ if absent. For $\phi=0$, $\pi=\beta$, Eq.~\eqref{eq:master} reduces to the stochastic NG of Ref.~\cite{baronchelli2007nonequilibrium}; for $m=2$ it closes on the three densities $x=n_{\{A\}}$, $y=n_{\{B\}}$, $z=n_{\{A,B\}}$ and yields Eqs.~\eqref{eq:mfx}--\eqref{eq:mfy}. For general $m$ the hierarchy is $(2^m{-}1)$-dimensional and not solvable in closed form.

A tractable closure is obtained by assuming that all active words are
statistically equivalent, so that $n_S$ depends on $S$ only through
its size: $n_S = \rho_k/\binom{m}{k}$ for $|S|=k$, where
$\rho_k(t)=\sum_{|S|=k} n_S(t)$ is the inventory-size distribution.
This closure is the inventory analogue of the heterogeneous
(degree-based) mean-field approach to dynamical processes on complex
networks~\cite{castellano2009statistical}, with the inventory size $k$
playing the role of the node degree. Under this ansatz, a listener
with inventory size $k$ contains the transmitted word with probability
$h_k = k/m$, answers YES with probability
\begin{equation}
  Q_k \;=\; \pi\,h_k + \phi\,(1-h_k) \;=\; \phi + (\pi-\phi)\,\frac{k}{m}\,,
  \label{eq:Qk}
\end{equation}
and learns a new word (out-inventory NO event) with probability
\begin{equation}
  L_k \;=\; (1-\phi)\Bigl(1-\frac{k}{m}\Bigr).
  \label{eq:Lk}
\end{equation}
Writing $Q=\sum_k \rho_k Q_k = \phi + (\pi-\phi)\,\mu/m$ for the
listener-averaged YES probability, with $\mu=\sum_k k\rho_k$ the mean
inventory size, the closed equations read
\begin{align}
  \dot\rho_1 &= Q\,(1-\rho_1) + \sum_{\ell\ge 2}\rho_\ell\, Q_\ell
               - \rho_1 L_1\,,
  \label{eq:rho1}\\
  \dot\rho_k &= -\rho_k\,(Q + Q_k) + \rho_{k-1}L_{k-1} - \rho_k L_k\,,
  \qquad 2\le k\le m.
  \label{eq:rhok}
\end{align}
In Eq.~\eqref{eq:rhok}, $-\rho_k Q$ is the loss of size-$k$ speakers
that collapse to singletons after a YES interaction, $-\rho_k Q_k$ the
loss of size-$k$ listeners that collapse, $\rho_{k-1}L_{k-1}$ the gain
of size-$k$ listeners by learning one new word, and $-\rho_k L_k$ the
corresponding loss towards size $k{+}1$.

At stationarity, Eq.~\eqref{eq:rhok} gives the recursion
\begin{align}
  \rho_k &= \rho_{k-1}\,\frac{L_{k-1}}{Q + Q_k + L_k}\,,
  \label{eq:stationaryrec}\\
  \text{i.e.}\quad
  \rho_k &= \rho_1 \prod_{r=1}^{k-1}\frac{L_r}{Q + Q_{r+1} + L_{r+1}}\,,
  \label{eq:stationary}
\end{align}
subject to the normalization $\sum_k \rho_k = 1$ and the
self-consistency condition $Q = \sum_k \rho_k Q_k$. The
large-vocabulary mean-field problem thus reduces to a scalar
self-consistency equation for $Q$, from which all $\rho_k$ follow
recursively. This word-exchangeable ansatz describes the symmetric
active phase; the ordering instability that breaks it is governed by
the two-word sector, Eq.~\eqref{eq:ordering}. The analysis of the
possible fragmented states of this hierarchy is left to future work.

\bibliographystyle{apsrev4-2}
\bibliography{refs}

\end{document}